\documentclass[showpacs,amsmath,amssymb,twocolumn,prx,superscriptaddress]{revtex4-1}

\usepackage{qcircuit}
\usepackage[dvips]{graphicx}
\usepackage{siunitx}
\usepackage{amsmath,amssymb,amsthm,mathrsfs,amsfonts,dsfont}
\usepackage{subfigure, epsfig}
\usepackage{braket}
\usepackage{bm}
\usepackage{enumerate}
\usepackage[dvipsnames]{xcolor}
\usepackage{color}
\usepackage{fancybox, graphicx}
\usepackage[skins]{tcolorbox}
\usepackage{multirow}
\usepackage{mathtools}

\usepackage[colorlinks]{hyperref}
\hypersetup{
	colorlinks=true,
	citecolor = {blue},
    linkcolor=blue,
    filecolor=magenta,      
    urlcolor=cyan,
}

\newtcolorbox[auto counter]{tbox}[2][]{%
	enhanced, float=hbt, drop fuzzy shadow southeast,
	colback=white!5!white, colframe=white!30!black,
	width= .97\columnwidth,sharp corners,boxrule=0.8pt,
	title={#2}, #1
}

\newtcolorbox{codebox}{enhanced, width=.95\columnwidth, halign = flush left, drop fuzzy shadow southeast, boxrule=0.4pt, sharp corners, colframe=black, colback=white}
\begin{document}

\title{Improving the accuracy of quantum computational chemistry using the transcorrelated method}

\begin{abstract}
Accurately treating electron correlation in the wavefunction is a key challenge for both classical and quantum computational chemistry. Classical methods have been developed which explicitly account for this correlation by incorporating inter-electronic distances into the wavefunction. The transcorrelated method transfers this explicit correlation from the wavefunction to a transformed, non-Hermitian Hamiltonian, whose right-hand eigenvectors become easier to obtain than those of the original Hamiltonian. In this work, we show that the transcorrelated method can reduce the resources required to obtain accurate energies from electronic structure calculations on quantum computers. We overcome the limitations introduced by the non-Hermitian Hamiltonian by using quantum algorithms for imaginary time evolution. 
\end{abstract}

\date{\today}
\author{Sam McArdle}
\email{sam.mcardle.science@gmail.com}
\affiliation{Department of Materials, University of Oxford, Parks Road, Oxford OX1 3PH, United Kingdom}

\author{David P. Tew}
\affiliation{Physical and Theoretical Chemical Laboratory, University of Oxford, South Parks Road, Oxford OX1 3QZ, United Kingdom}

\maketitle

\section{Introduction}\label{Sec:Intro}
Since its conception, scientists have applied the theory of quantum mechanics to better understand the properties of atoms, molecules, and materials. Despite considerable progress in this area, simulating realistic systems to high accuracy remains a central challenge for computational chemistry. This difficulty stems from the computational complexity of exactly simulating many-body quantum systems; a problem believed to require classical resources scaling exponentially with the size of the simulated system. It is believed that using a controllable quantum system as the simulation platform (as originally proposed by Feynman~\cite{feynman1982simulating}) may ameliorate this problem. Quantum algorithms have been developed that require polynomially scaling resources to obtain both the static and dynamic properties of quantum systems of interest (see Refs.~\cite{cao2018chemreview, mcardle2020review, bauer2020QchemReview} and the references therein). 

Nevertheless, despite the optimism around the growing field of quantum computational chemistry, many challenges still remain. Most notably, current quantum computers are limited in qubit count and circuit depth, and so cannot yet outperform classical computers at useful tasks. Herein, we focus on the problem of determining the electronic structure of molecules and condensed matter systems.

Wavefunctions describing the ground and excited states of these systems need to account for the correlation between electrons. 
Electron correlation is loosely divided into `static' and `dynamic' correlation. Static correlation refers to cases where more than one Slater determinant is dominant
in the wavefunction, such as during bond breaking, or low-spin open-shell states. Static correlation is typically dealt with using non-perturbative techniques that seek to accurately treat the dominant components of the wavefunction, such as multiconfigurational self-consistent field theory (MCSCF), and tensor network methods. Dynamic correlation refers to the minor fluctuations from a mean field resulting from instantaneous Coulomb repulsion between electrons, which tends to anti-correlate their positions. Dynamic correlation is usually dealt with using a coupled cluster expansion or perturbation theory. The most severe manifestation of dynamic correlation occurs at short inter-electronic distances. As two electrons approach each other, the Coulomb potential diverges, leading to sharp features in the wavefunction, known as the `electron-electron cusps'. Accurately resolving these cusps typically involves expanding the wavefunction as a linear combination of Slater determinants in a large single particle basis set. This is often the approach taken in coupled cluster, or perturbation theory calculations. 

It has proven challenging to develop techniques that can accurately treat both static and dynamic correlation, at reasonable cost. The former techniques are too expensive to apply in large basis sets, while the latter techniques are typically designed to work from single determinant reference states. Although techniques have been developed to resolve this issue, such as MCSCF + perturbation theory, there is not yet a clear consensus as to which methods are most effective. This problem even extends to simulations performed on quantum computers. While quantum computers can easily account for static correlation by efficiently storing the many-body wavefunction, increasing the basis set size and the number of excitations considered requires additional qubits and gates. As current quantum computers are limited in both qubit count and gate depth, this restricts the size and accuracy of the quantum simulations that we can carry out. \\

A number of methods have been developed which target dynamic correlation, while using smaller basis sets. These techniques typically introduce an explicit dependence on the electron-electron distances into the wavefunction. The leading techniques are reviewed in Refs.~\cite{gruneis2017correlatedperspective, hattig2012correlatedreview,Kong2012explicitlycorrelated}, and include: Hylleraas-type methods, exponentially correlated Gaussians, the transcorrelated method, and R12/F12 methods. In this work, we focus on the transcorrelated method of \textcite{boys1969tc1, boys1969tc2}. This method does not depend on the algorithm used to solve the resulting chemistry problem, and does not include approximations. This motivates replacing the classical chemistry methods typically used for the transcorrelated approach, with a quantum subroutine. However, as the transcorrelated Hamiltonian is no longer Hermitian, it is not straightforward to incorporate it into canonical quantum algorithms like the variational quantum eigensolver (which relies on having a variational lower bound when measuring the expectation values of Hermitian operators), or quantum phase estimation (which evolves the system under a unitary function of the Hamiltonian). Instead, we make use of a quantum algorithm implementing ansatz-based imaginary time evolution~\cite{mcardle2018variational}, motivated by the success of imaginary time-like methods such as full configuration interaction quantum Monte Carlo (FCIQMC) when used in conjunction with the transcorrelated method~\cite{luo2018TC_FCIQMC_planewave,dobrautz2019TC_hubbard, cohen2019TCatoms}. \\

The rest of this paper proceeds as follows. In Sec.~\ref{Sec:TC_approach} we introduce explicitly correlated methods, discuss the transcorrelated method in detail, and highlight some of the basic properties of non-Hermitian operators. In Sec.~\ref{Sec:ImagTime_TC} we show that quantum imaginary time evolution algorithms can in principle be used to find the ground states of non-Hermitian Hamiltonians, including the transcorrelated Hamiltonian. We confirm this numerically in Sec.~\ref{Sec:Results}, where we simulate using our method to find the ground state of small Fermi-Hubbard models to high accuracy.

We note that while this manuscript was being finalised, a related preprint was released~\cite{motta2020TC}. That work considers the `canonical transcorrelated' approach of \textcite{yanai2012canonicalTC}, which leads to a unitary transformation of the Hamiltonian, and thus a Hermitian transcorrelated Hamiltonian. This unitarity comes at a cost of introducing approximations in the Hamiltonian, which result from truncating the Baker–Campbell–Hausdorff (BCH) expansion of the transformed Hamiltonian at second order. The severity of this approximation is not yet fully understood by the computational chemistry community. Nevertheless, the Hermitian nature of the (approximate) canonical transcorrelated Hamiltonian enables it to be used in black-box quantum algorithms like the variational quantum eigensolver (VQE) and quantum phase estimation (QPE). The authors of that work use the VQE to obtain more accurate results than would be obtained by using the unmodified Hamiltonian. As such, the results of our two papers are complementary, and highlight the value in further exploring transcorrelated approaches to quantum computational chemistry.

\section{Explicitly correlated methods}\label{Sec:TC_approach}

Before discussing explicitly correlated approaches to quantum chemistry, we first briefly review some of the standard approaches used to solve the electronic structure problem. A comprehensive discussion of these techniques is given in the textbooks by \textcite{helgaker2014molecular,szabo2012modern}. We are often interested in obtaining the low lying eigenstates of chemical systems of interest, as these will determine their chemical properties. As a first approximation, we typically restrict ourselves to solving the electronic Schr\"odinger equation, which is obtained by fixing the positions of the nuclei, and treating them as classical particles (the Born-Oppenheimer approximation). We seek the eigenvalues and eigenstates of the electronic Hamiltonian, which in atomic units is
\begin{equation}\label{ElectronicStructureH}
	H_e = -\sum_i\frac{\nabla^2_i}{2} -\sum_{i,I}\frac{Z_I}{|\mathbf{r}_i-\mathbf{R}_I|}+\frac{1}{2}\sum_{i\neq j}\frac{1}{|\mathbf{r}_i-\mathbf{r}_j|},
\end{equation}
where $\mathbf{R}_I$, and $Z_I$ denote the position, and atomic number of the $I$\textsuperscript{th} nucleus, and $\mathbf{r}_i$ is the position of the $i$\textsuperscript{th} electron. This problem is too difficult to solve for all but the simplest systems, such as the hydrogen atom. To make the calculation more tractable, the continuous, real space Hamiltonian is usually projected onto a finite Hilbert space defined by a basis set, typically a set of $M$ atomic or molecular spin-orbitals. The vector space is then the set of all possible anti-symmetrised products (Slater determinants) of $N$ electrons in $N$ spin-orbitals. The anti-symmetry requirement, dictated by the Pauli principal, is naturally taken care of in the language of second quantisation. In second quantisation, we can write the wavefunction as 
\begin{equation}\label{2ndQwavefunc}
	\begin{aligned}
		\ket{\Psi} = \sum_i \alpha_i \ket{i} ,
	\end{aligned}
\end{equation}
where $\alpha_i$ are complex coefficients, and $\ket{i}$ represent Slater determinants, which are conveniently represented by the shorthand notation of Fock occupation number vectors
\begin{equation}\label{Slater}
    \begin{aligned}
    \ket{i} = \ket{i_{M-1}, \dots, i_j,\dots, i_0},
    \end{aligned}
\end{equation}
where $i_j = 1$ when spin-orbital $j$ is occupied in the corresponding Slater determinant, and $i_j = 0$ when it is empty. The electronic Hamiltonian projected onto a basis of single particle orbitals can be written as
\begin{equation}\label{FH}
	H = \sum_{p,q}h_{pq}a^\dag_p a_q + \frac{1}{2}\sum_{p,q,r,s}h_{pqrs}a^\dag_p a^\dag_q a_ra_s,
\end{equation}
where $h_{pq}$ and $h_{pqrs}$ are coefficients obtained from single and two-particle integrals (respectively), and $a^\dag_i$, $a_i$ are fermionic creation and annihilation operators (respectively). 

Many of the standard methods in computational chemistry, such as: the Hartree--Fock, coupled cluster, and configuration interaction methods, and multiconfigurational self-consistent field theory, all consider wavefunctions of the form given by Eq.~(\ref{2ndQwavefunc}). They seek to approximate the full configuration interaction (FCI) solution to the electronic Schr\"odinger equation by either including a limited number of the determinants in the expansion, or by including all of the terms, but with an approximation to the true $\alpha_i$ values. Unfortunately, it was realised as early as the 1920's that expanding the wavefunction as a linear combination of Slater determinants leads to a slow convergence to the true eigenvalues and eigenvectors of Eq.~(\ref{ElectronicStructureH}). This is due 
to the failure of the basis expansion to resolve the sharp features in the many-body wavefunction at electron-electron coalescence.
The `cusp conditions' that the wavefunction must obey to accurately represent the true system were formalised by \textcite{kato1957cusp} -- wavefunctions constructed from products of single particle orbitals do not fulfil these conditions~\cite{nooijen1998InfinityElimination}.\\

However, long before Kato's mathematically rigorous description of the cusp conditions, it was known that the convergence of quantum chemistry calculations could be accelerated by explicitly including functions of the inter-electronic distances in the wavefunction. The first of these calculations were carried out by \textcite{Hylleraas1929helium, slater1928rydberg}. These methods give extremely accurate results for small systems, but are restricted in their applicability by their need to evaluate $N$-electron integrals, where $N$ is the number of electrons in the system. A number of methods have since been developed to make calculations of this type more practical. These are collectively known as explicitly correlated methods, and have been reviewed by \textcite{hattig2012correlatedreview,gruneis2017correlatedperspective,Kong2012explicitlycorrelated}. As the main focus of this work is the transcorrelated method, we will only reference the other main explicitly correlated approaches, before discussing in detail the transcorrelated method. 

The `exponentially correlated Gaussians' approach is similar in spirit to the Hylleraas method. The method considers wavefunctions that depend explicitly on the distances between electrons. This approach is variational, and can obtain extremely accurate results for molecules with up to around 3 electrons, or atoms with 4-6 electrons~\cite{hattig2012correlatedreview}. However, it is again constrained by the requirement to carry out $N!$ $N$-electron integrals. The exponentially correlated Gaussians approach can be simplified to the Gaussian geminals method, which considers basis functions restricted to the coordinates of two electrons. This approach can be combined with pair theories, such as M\o ller--Plesset second order perturbation theory (MP2) or coupled cluster with single and double excitations (CCSD). 

A more practical approach is the R12 method, and its modern variant, termed F12. This technique was originally introduced by \textcite{Kutzelnigg1985r12}, and augments a standard single particle basis expansion with excitations of pairs of electrons into two-electron basis functions with a specific form. F12 methods consider a more general form for the two-electron functions than the original R12 methods. F12 methods have proven effective at accurately treating large systems. They are used in conjunction with pair theories like CCSD and MP2, and CCSD-F12 methods in small basis sets can be used to obtain results as accurate as CCSD at the basis set limit~\cite{hattig2012correlatedreview}. F12 methods can be considered the leading explicitly correlated approach. However, their efficient implementation utilises a number of approximations that can complicate calculations.\\

An alternative approach to explicitly dealing with dynamic correlation in the wavefunction is the transcorrelated (TC) method. The TC method was introduced by \textcite{boys1969tc1, boys1969tc2}, and can be linked back to an earlier approach by \textcite{hirschfelder1963similarity}. These works observed that rather than considering the wavefunction to be transformed by an auxiliary function that describes dynamic correlation, it is equivalent to consider the Hamiltonian to be transformed by the auxiliary function. This can be likened to working in the Heisenberg picture of quantum mechanics (where operators are made time-dependent, and wavefunctions are time-independent) rather than the Schr\"odinger picture (where wavefunctions change in time, and operators are unchanging). To apply the TC method, we can write that
\begin{equation}\label{Eq:TC_wf}
    \ket{\psi} = e^{\sum_{i<j} f(\textbf{r}_i, \textbf{r}_j)} \ket{\phi} = e^{\hat{g}} \ket{\phi}
\end{equation}
where $\ket{\psi}$ is the wavefunction of the system, $f(\textbf{r}_i, \textbf{r}_j)$ is a symmetric, real function of the positions of electrons $i$ and $j$ (referred to as a Jastrow factor), and $\ket{\phi}$ is a wavefunction that does not explicitly depend on inter-electronic distances. Considering the solutions of the real space electronic structure Hamiltonian, we see that 
\begin{align}\label{Eq:TC_energy_rhs}
    H\ket{\psi_i} &= E_i\ket{\psi_i} \\
    \rightarrow H e^{\hat{g}} \ket{\phi_i} &= E_i e^{\hat{g}} \ket{\phi_i} \\
    \rightarrow H' \ket{\phi_i} &= E_i \ket{\phi_i},
\end{align}
where $H$ is the Hamiltonian of the system, and $H' = e^{-\hat{g}} H e^{\hat{g}}$ is defined as the transcorrelated Hamiltonian. While the explicitly correlated wavefunction $\ket{\psi_i}$ is an eigenstate of the original Hamiltonian, we see that we can obtain the same eigenvalue by finding the wavefunction $\ket{\phi_i}$, which is an eigenstate of $H'$. As $\ket{\phi_i}$ is not explicitly correlated, it should be easier to obtain than $\ket{\psi_i}$. As the transformation $e^{\hat{g}}$ is not unitary, the TC Hamiltonian $H'$ is no longer Hermitian. This leads to a number of issues, including: a lack of variational lower bound on the ground state eigenvalue, different right-hand and left-hand eigenvectors, and non-orthogonal right-hand (or left-hand) eigenvectors.\\

Since the TC transformation is performed prior to projection onto a basis set, the projected TC Hamiltonian is not isospectral with the original projected Hamiltonian, except in the limit of an infinite basis set. The Jastrow factor is chosen to regularise the Hamiltonian, such that the TC Hamiltonian is free from singularities, and to ensure that the function $e^{\hat{g}} \ket{\phi_i}$ has the correct behaviour in the region of the electron-electron and electron-nucleus coincidences. If both the TC Hamiltonian and the unmodified Hamiltonian are projected onto the same single particle basis set, the TC Hamiltonian will yield energies closer to those obtained in the basis set limit. The TC Hamiltonian in the real space formulation is obtained from the BCH expansion of $H' = e^{-\hat{g}} H e^{\hat{g}}$, which truncates at second order in $g$~\cite{cohen2019TCatoms}
\begin{align}\label{Eq:TC_BCH}
    H' &= H + [H, \hat{g}] + \frac{1}{2} [[H, \hat{g}],\hat{g}] \\
    &= H - \sum_i \bigg{(} \frac{1}{2} \nabla_i^2 \hat{g} +(\nabla_i \hat{g}) \nabla_i + \frac{1}{2} (\nabla_i \hat{g})^2 \bigg{)}.
\end{align}
This leads to additional two and three-body terms in the Hamiltonian. The TC transformation may also provide additional benefits beyond effectively expanding the size of the basis set used. For example, one can also apply a TC-type transformation to the Hamiltonian after it is projected onto single particle basis functions. This was applied to the Fermi-Hubbard model by \textcite{tsuneyuki2008TChubbard,dobrautz2019TC_hubbard}. They considered transformation with a Gutzwiller factor
\begin{equation}\label{Eq:TC_hubbard_1}
    H' = \bigg{(}e^{-J\sum_i n_{i, \uparrow} n_{i, \downarrow} } \bigg{)} H \bigg{(} e^{J\sum_j n_{j, \uparrow} n_{j, \downarrow} } \bigg{)},
\end{equation}
where $n_{i,\sigma}$ is the number operator for the spin-lattice site indexed by $i, \sigma$. The transformation acts to suppress double occupancies of lattice sites~\cite{dobrautz2019TC_hubbard}. While the resulting TC Hamiltonian is still isospectral to the unmodified Hamiltonian in this case, those authors observed other benefits introduced by the TC method. They found that the TC Hamiltonian had more `compact' right-hand eigenvectors than the regular Hamiltonian, which made it easier to approximate them to high accuracy. This `compactification' of right-hand eigenvectors also persists when the TC method is applied before projecting onto single particle basis functions, as described above~\cite{cohen2019TCatoms}. 

Unfortunately, the benefits of the TC method may be considered a double-edged sword; while the right-hand eigenvectors are made easier to obtain, the left-hand eigenvectors gain additional dynamic correlation, and thus may become more difficult to construct from a single particle basis expansion~\cite{nooijen1998InfinityElimination}. The left-hand eigenvector is given by
\begin{align}\label{Eq:LeftEig}
    \bra{\psi_i}H &= \bra{\psi_i}E_i \\
    \rightarrow \bra{\phi_i} e^{\hat{g}} H &= \bra{\phi_i}e^{\hat{g}} E_i \\
    \rightarrow \bra{\phi_i}e^{2\hat{g}} H' &= \bra{\phi_i}e^{2\hat{g}} E_i, \\
    \bra{\tilde{\phi}_i}H' &= \bra{\tilde{\phi}_i} E_i
\end{align}
where $\bra{\tilde{\phi}_i} = \bra{\phi_i}e^{2\hat{g}}$ is the left-hand eigenvector of $H'$. The differing forms of the left-hand and right-hand eigenvectors prove problematic for measuring observables other than the energy. For example, we see that
\begin{align}\label{Eq:Observable}
    \langle O \rangle = \bra{\psi} \hat{O} \ket{\psi} = \bra{\phi} e^{\hat{g}} \hat{O} e^{\hat{g}} \ket{\phi}.
\end{align}\\
As $e^{\hat{g}} \hat{O} e^{\hat{g}}$ does not have a terminating BCH expansion, we use the expansion $\hat{O}' = e^{-\hat{g}} \hat{O} e^{\hat{g}}$ (which does truncate) to write that 
\begin{equation}\label{Eq:Observable}
    \langle O \rangle = \bra{\phi} e^{2\hat{g}} \hat{O}' \ket{\phi} = \bra{\tilde{\phi}_i} \hat{O}' \ket{\phi}.
\end{equation}
As a result, calculating observables other than the energy requires obtaining the left-hand eigenvector of the TC Hamiltonian, which the TC transformation makes more difficult to obtain. This challenge has yet to be resolved in studies on the TC method, which have mainly focused on finding the ground state energy of various systems. \\

Early works on the TC method considered a reference state $\ket{\phi}$ consisting of a single Slater determinant, and optimised the parameters in the Jastrow function and the form of the single-particle orbitals using self-consistent equations (known as the TC-SCF method)~\cite{boys1969tc1, boys1969tc2}. However, the use of the non-Hermitian TC Hamiltonian means that it is not possible to use the Rayleigh-Ritz variational principle to minimise the energy~\cite{handy1971TranVariance}. This makes it difficult to confirm that a good solution has been found, even if the TC equations appear converged. \textcite{handy1971TranVariance} suggested minimising the variance of the energy to find the ground state. However, this approach presents two issues. Firstly, minimising the variance in variational calculations typically gives a less accurate energy estimate than minimising the energy itself~\cite{handy1971TranVariance}. Secondly, the TC Hamiltonian $H'$ projected onto a single particle basis contains up to $O(M^6)$ terms, compared to $O(M^4)$ in the unmodified Hamiltonian, where $M$ is the number of spin-orbitals included in the single-particle basis set. This is due to the inclusion of 3-electron terms in the TC Hamiltonian. These computational difficulties prevented the TC method from gaining widespread use.

More recent work has attempted to make the TC method more practical. \textcite{tenno2000transcorrelated, hino2002transCCSD} fixed the form of the Jastrow function, and compensated for the error this introduces by expanding $\ket{\phi}$ as a sum of Slater determinants. An alternative approach combined the TC-SCF method with variational Monte Carlo applied to the energy variance, to construct an iterative optimisation procedure~\cite{umezawa2003TCmontecarlo, umezawa2004excitedTC,umezawa2005threebody,tsuneyuki2008TChubbard,ochi2012TCperiodic}. \textcite{luo2010variationalTCthrowaway,luo2011variationalTCmcscf} constructed a variational TC-SCF approach by discarding the terms in the TC Hamiltonian linear in $g$ to obtain a Hermitian operator.

As can be seen above, a key challenge to overcome has been how to best optimise the wavefunction, given the non-variational nature of the energy, and the difficulties with optimising parametrised Jastrow factors. \textcite{luo2018TC_FCIQMC_planewave} developed an approach to overcome these limitations. They used a wavefunction comprised of a frozen Jastrow term, and a Slater determinant expansion, where the determinant expansion was optimised using full configuration interaction quantum Monte Carlo (FCIQMC)~\cite{booth2009FCIQMCoriginal}. FCIQMC is closely related to imaginary time evolution of the state, and should converge to the ground state if a sufficiently long time evolution is used. This method can be used to find the ground state of the TC Hamiltonian, without invoking variational properties
\begin{align}\label{Eq:ImagTime_unnorm}
    \ket{\psi_0} &= \lim_{\tau\rightarrow\infty} e^{-H\tau}\ket{\psi}, \\
    \rightarrow     e^{\hat{g}} \ket{\phi_0} &= \lim_{\tau\rightarrow\infty} \sum_{k=0}^\infty \frac{(e^{\hat{g}} H'e^{-\hat{g}}\tau)^k}{k!} e^{\hat{g}} \ket{\phi}, \\
    \rightarrow     \ket{\phi_0} &= \lim_{\tau\rightarrow\infty} e^{-H'\tau}\ket{\phi}.
\end{align}
By transferring the dynamic correlation from the right-side wavefunction to the TC Hamiltonian, \textcite{luo2018TC_FCIQMC_planewave} were able to make the Slater determinant wavefunction expansion more compact, which is beneficial for the FCIQMC method since far fewer walkers are required to accurately sample the wavefunction expansion. This approach has been applied to simulations of plane wave Hamiltonians~\cite{luo2018TC_FCIQMC_planewave}, the Fermi-Hubbard model~\cite{dobrautz2019TC_hubbard}, atomic systems in Gaussian basis sets~\cite{cohen2019TCatoms}, quantum gases~\cite{jeszenszki2018TCgases}, and ultracold atoms~\cite{jeszenszki2020TCultracold}. \\

Given the relative simplicity of the TC method (compared to methods such as F12), and the fact that it is agnostic of the approach used to generate the Slater determinant expansion for $\ket{\phi_i}$, it is a natural target for incorporation into algorithms which use quantum computers to solve the electronic structure problem.  However, the non-Hermitian nature of the TC Hamiltonian presents a significant challenge to overcome in this regard. The two main quantum algorithms for solving the electronic structure problem are the variational quantum eigensolver (VQE)~\cite{peruzzo2014variational,VQETheoryNJP}, and quantum phase estimation~\cite{kitaev1995phase}. These algorithms are thoroughly reviewed in Refs.~\cite{mcardle2020review, cao2018chemreview,bauer2020QchemReview}, but we summarise the key details here. 

The VQE uses parameterised quantum circuits to generate ansatz states for the eigenstates of interest. The energy of the ansatz state can be measured, and is then input (as classical data) into a classical optimisation algorithm, together with the current parameters of the quantum circuit. The optimisation algorithm then outputs new parameters, which should yield a lower energy state. This procedure is iterated until the energy converges, ideally to the ground state of the system (relying upon the Rayleigh-Ritz variational principle). As the qubits are measured after each construction of the ansatz state, the circuit depth of the VQE may be kept relatively shallow. As such, it is hoped that the ansatz states can be constructed before noise is able to build up, enabling the algorithm to proceed without quantum error correction. The main limitation of the VQE is the use of short circuits to lessen the effects of noise, which in turn limits the quality of the approximation to the ground state that we can obtain. As the VQE utilises the variational lower bound on the energy, resulting from the Hermitian nature of the Hamiltonian, it appears difficult to integrate with the non-Hermitian TC Hamiltonian. An obvious solution is to minimise the variance of the TC Hamiltonian. However, as discussed above, the energies obtained from variance minimisation are typically less accurate than energies obtained from direct energy minimisation. In addition, the number of measurements would be on the order of $O(M^{12})$ for molecular systems in a Gaussian basis set, which would quickly become infeasible.  

In contrast, QPE (in its canonical form) proceeds by coherently evolving the system under a unitary, isospectral, and invertible function of the Hamiltonian, controlled upon the state of an ancillary register. This accrues an energy dependent phase on the ancillary register, which can be measured through an inverse quantum Fourier transform. Measuring the ancillary register to obtain the energy eigenvalue projects the main register into the corresponding FCI eigenstate. QPE succeeds with high probability if the unitary evolution is carried out for a sufficiently long duration, and if the initial state of the main register has a non-negligible overlap with the FCI ground state. Due to the long coherent evolutions required by QPE, it is typically considered to require quantum error correction, and thus is not considered a near-term approach. The required unitary evolution under an isospectral, and invertible function of the Hamiltonian appears difficult to achieve with the non-Hermitian TC Hamiltonian, again suggesting that the two methods may not be compatible. 

To the best of our knowledge, only a few works have considered problems closely related to this area. As discussed in the introduction, the recent work of \textcite{motta2020TC} investigated the use of an approximate (Hermitian) TC Hamiltonian, which is therefore compatible with the VQE and QPE. \textcite{bauman2019downfolding, bauman2019downfoldingexcited} developed a unitary transformation based on unitary coupled cluster theory (UCC), that `downfolds' some of the dynamic correlation of the system into a smaller active space. The resulting Hamiltonian is Hermitian, and is obtained through classical precomputation, involving CC or UCC calculations on the system. A related method was introduced by \textcite{takeshita2019virtualorbs}, which considers double excitations from active orbitals into virtual orbitals. They show that the effect of these excitations can be incorporated into the ground state energy estimate using only measurements on the qubits representing active space orbitals. This means that the virtual orbitals do not need to be included in the simulation, thus reducing the number of qubits required. \\

Motivated by the success of FCIQMC at dealing with the TC Hamiltonian (see Eq.~(\ref{Eq:ImagTime_unnorm})), we consider quantum algorithms for imaginary time evolution. Two possibilities exist; ansatz-based quantum imaginary time evolution~\cite{mcardle2018variational}, and Trotter-based quantum imaginary time evolution~\cite{motta2019imaginary}. The latter method would require a large circuit depth, owing to the requirement of dividing evolution under the TC Hamiltonian, which contains up to $O(M^6)$ terms, into a number of Trotter steps. As such, in this work we focus on applying ansatz-based quantum imaginary time evolution to find the ground state of the TC Hamiltonian.

\section{Ansatz-based quantum imaginary time evolution}\label{Sec:ImagTime_TC}

Imaginary time evolution is a powerful method for finding the ground states of quantum systems - even when performed on classical computers, and constrained to a manifold of states generated from an ansatz~\cite{hackl2020GeometryImagTime}. Following the development of an ansatz-based quantum algorithm for real time evolution~\cite{Li2017}, a related quantum algorithm carrying out ansatz-based imaginary time evolution was developed~\cite{mcardle2018variational}. These ansatz-based time evolution algorithms were formalised by \textcite{yuan2018variationaltheory}, and later extended to the simulation of general processes~\cite{endo2018variational}, and mixed states~\cite{koczor2019NaturalGrad}. The original ansatz-based quantum imaginary time evolution work~\cite{mcardle2018variational} applied the method to find the ground states of small molecular systems, in minimal basis sets. The algorithm was found to adhere closely to the true imaginary time dynamics. The algorithm has since been successfully applied to: finding excited states of quantum systems~\cite{jones2019ImagExcited}, recompiling~\cite{jones2018compiling} and discovering~\cite{xu2019compiling} quantum circuits, training quantum Boltzmann machines for quantum machine learning~\cite{zoufal2020ImagTimeBoltzmann}, simulating models of quantum field theories~\cite{liu2020ImagTimeFieldTheories}, solving systems of linear equations~\cite{xu2019LinearAlgebra, huang2019LinearAlgebra}, and pricing financial options~\cite{fontanela2019ImagTimeFinance}.\\

The algorithm proceeds by using a parameterised quantum circuit (an `ansatz'), $U(\vec{\theta}_\tau)$, to represent the state of the quantum system at a given point $\tau$ on its imaginary time trajectory. To evolve the state forward in imaginary time, the circuit parameters are updated according to an update rule derived from McLachlan's varitional principle~\cite{mclachlan1964variational} (although one can also derive the algorithm from the Dirac and Frenkel variational principles~\cite{yuan2018variationaltheory}) applied to the the imaginary time Schr\"odinger equation. The algorithm thus seeks the state that is closest in distance to the state obtained from `true' imaginary time evolution, but that can still be prepared by the ansatz circuit. The minimisation of an energy cost function thus happens as a corollary of a sufficiently long propagation in imaginary time, rather than due to an application of the Rayleigh-Ritz variational principle. As such, we expect that the method should be able to converge to the ground state of the non-Hermitian TC Hamiltonian, as suggested by Eq.~(\ref{Eq:ImagTime_unnorm}). 

As we will construct the ansatz state, which we denote as $\ket{\phi(\vec{\theta}_\tau)} = \ket{\phi(\tau)}$, on a quantum computer, it must be normalised. We therefore seek to propagate the initial state $\ket{\phi(0)}$ in imaginary time
\begin{equation}\label{Eq:Imag_time_ansatz}
    \ket{\phi(\tau)} = \frac{e^{-H'\tau}\ket{\phi(0)} }{\sqrt{\bra{\phi(0)}e^{-H'^\dag\tau} e^{-H'\tau}\ket{\phi(0)} }},
\end{equation}
with $H' = e^{-\hat{g}} H e^{\hat{g}}$. This corresponds to imaginary time evolution of an un-normalized state
\begin{equation}
    \ket{\psi(\tau)} = \frac{e^{-H\tau}\ket{\psi(0)} }{\sqrt{\bra{\psi(0)}e^{-H\tau} e^{-2\hat{g}} e^{-H\tau}\ket{\psi(0)} }},
\end{equation}
which obeys the normalisation condition that we would expect. We can verify that the state given by Eq.~(\ref{Eq:Imag_time_ansatz}) satisfies a modified version of the imaginary time Schr\"odinger equation
\begin{equation}\label{Eq:Imag_time_SE}
    \frac{\partial \ket{\phi(\tau)}}{\partial \tau} = -[H' - \Re(E_\tau)]\ket{\phi(\tau)},
\end{equation}
where $\Re(E_\tau)$ is the real part of $E_\tau = \braket{{\phi(\tau)}|H'|{\phi(\tau)}}$. McLachlan's variational principle applied to  Eq.~(\ref{Eq:Imag_time_SE}), is given by
 \begin{equation}
 	\delta \|({\partial}/{\partial \tau} + H'-\Re(E_\tau))\ket{\phi(\tau)}\|=0
 \end{equation}
where
\begin{equation}
\|\hat{A}\ket{\alpha}\|=\left(\hat{A}\ket{\alpha} \right)^\dag \left(\hat{A}\ket{\alpha}\right),
\end{equation}
for an arbitrary operator $\hat{A}$ and arbitrary state $\ket{\alpha}$. We show in the Appendix that by constraining $\ket{\phi(\tau)}$ to be constructed from a parameterised unitary quantum circuit, $\ket{\phi(\tau)} = U(\vec{\theta}_\tau) \ket{\bar{0}}$, we can obtain the following parameter update rule, which evolves the parameters such that the state generated propagates forwards in imaginary time
\begin{equation}\label{Eq:Imag_time_Matrix_vec}
	\begin{aligned}
		\sum_j A_{ij} \dot{\theta}_j = - C_i,
	\end{aligned}
\end{equation}
with
\begin{equation}\label{Eq:Explain_M_C}
\begin{aligned}
	A_{ij} &= \Re\left(\frac{\partial \bra{\phi(\tau)}}{\partial \theta_i}\frac{\partial \ket{\phi(\tau)}}{\partial \theta_j}\right),\\
	C_i &= \Re\left(\frac{\partial \bra{\phi(\tau)}}{\partial \theta_i} H'\ket{\phi(\tau)}\right).
\end{aligned}
\end{equation}
We can evolve the parameters in imaginary time by using an Euler update rule
\begin{equation}\label{Euler}
	\begin{aligned}
	\vec{\theta}({\tau + \delta \tau}) &\simeq \vec{\theta}(\tau) +  \dot{\vec{\theta}}(\tau)\delta \tau= \vec{\theta}(\tau) - A^{-1}(\tau)\cdot \vec{C}(\tau)\delta \tau.
	\end{aligned}
\end{equation}
These equations are identical to those derived in Ref.~\cite{mcardle2018variational} for imaginary time evolution under a Hermitian Hamiltonian. We can obtain the left-hand eigenstate of $H'$ by replacing $H'$ with $H'^\dag$ in the equations above.

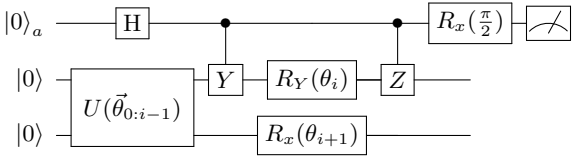
\begin{figure}[!h]
\begin{align*}
\Qcircuit @C=0.6em @R=.7em {
\lstick{\ket{0}_a}&\gate{\mathrm{H}}&\ctrl{1}&\qw&\ctrl{1}&\gate{R_x(\frac{\pi}{2})}&\meter\\
\lstick{\ket{0}}&\multigate{1}{U(\vec{\theta}_{0:i-1})}&\gate{Y}&\gate{R_Y(\theta_i)}&\gate{Z}\qw&\qw\\
\lstick{\ket{0}}&\ghost{U(\vec{\theta}_{0:i-1})}&\qw&\gate{R_x(\theta_{i+1})}&\qw\qw&\qw\\}
\end{align*}
\caption{A quantum circuit to calculate the $C_i$ term in Eq.~(\ref{Eq:Explain_M_C}), when the Hamiltonian is given by $H=Z$. The terms $A_{ij}$ can be measured using a similar probe circuit, as discussed in Refs.~\cite{Li2017, mcardle2018variational}.}\label{Fig:Qgrad}
\end{figure}

The values of $A_{ij}$ and $C_i$ can be obtained using `Hadamard test' circuits, of the type shown in Fig.~\ref{Fig:Qgrad}. These circuits must be repeated many times in order to evaluate each term of the metric $A_{ij}$ matrix. When measuring the components of the gradient vector $\vec{C}$, the circuit must be repeated a number of times, for each term in the Hamiltonian. These are then linearly combined, to obtain the gradient component for the given parameter. Recent work~\cite{VanStraaten2020Measurement} has investigated optimising the measurement of these terms, when we are restricted to a finite number of circuit repetitions. It is also possible to measure the gradient vector without introducing an additional probe qubit~\cite{mitarai2018imaginaryindirect,schuld2019gradients}. 

Once the state has been evolved in imaginary time, we can measure the energy. As the TC Hamiltonian is non-Hermitian, measuring the energy on an arbitrary state may return a complex expectation value. However, if we are able to obtain a state close to the desired eigenstate, the resulting imaginary component of the energy should be small, as the eigenvalues of the TC Hamiltonian are still real. As will be discussed in Sec.~\ref{Sec:Results}, the TC Hamiltonian can be mapped to a weighted sum of tensor products of single qubit Pauli operators. The energy can then be measured as 
\begin{equation}\label{VQEenergy}
	\begin{aligned}
		E(\vec{\theta}_\tau) = \sum_i h_i \langle \phi (\vec{\theta}_\tau) | P_i | \phi (\vec{\theta}_\tau) \rangle,
	\end{aligned}
\end{equation}
where $P_i$ is a tensor product of single qubit Pauli operators, and $h_i$ is the corresponding coefficient.

\section{Results}\label{Sec:Results}
In order to verify that the ansatz-based quantum imaginary time evolution algorithm is compatible with the TC method, we conduct numerical simulations on the TC Fermi-Hubbard model. Numerical simulations were carried out in Cirq~\cite{google2020cirq}, a Python package for simulating intermediate-sized quantum computers. Rather than evaluating all of the different $A_{ij}$ and $C_i$ terms using the quantum circuits described in Fig.~\ref{Fig:Qgrad}, we use a first order finite difference approximation to obtain the gradient, with a finite difference step size of $10^{-10}$. The inversion of the metric matrix is performed using a generalised inverse of the matrix, obtained from a singular-value decomposition that includes all singular values above $10^{-6}$. A timestep of $0.01$ was used for the Euler update rule.

The Fermi-Hubbard (FH) model is often considered as a simplified model for high-temperature cuprate superconductors~\cite{dagotto1994correlatedelectrons,anderson2002hubbard}. The unmodified FH Hamiltonian is given by
\begin{equation}\label{Eq:Hubbard_unmodified}
H_{\mathrm{FH}}=-t \sum_{\langle i, j\rangle, \sigma}\left(a_{i, \sigma}^{\dagger} a_{j, \sigma}+a_{j, \sigma}^{\dagger} a_{i, \sigma}\right)+U \sum_{i} n_{i, \uparrow} n_{i, \downarrow}
\end{equation}
where $\langle i, j\rangle$ denotes a sum over nearest-neighbour lattice sites, and $\sigma$ is a spin-coordinate. This Hamiltonian describes fermions hopping between nearest-neighbour lattice sites with strength $t$, which are subject to a repulsive (or attractive) force $U$ when they occupy the same lattice site. The ground state of this Hamiltonian is difficult to obtain for large system sizes, at close to half-filling, and at interaction strengths in the region of $t=1$, $U=4$~\cite{simonscollab2015hubbard}. As such, we choose this interaction strength for our numerical simulations.

The TC Hamiltonian was obtained using the Gutzwiller transformation of \textcite{tsuneyuki2008TChubbard,dobrautz2019TC_hubbard}, as described in Sec.~\ref{Sec:TC_approach}. The TC Hamiltonian is given by
\begin{equation}\label{Eq:TC_hubbard_2}
    H_{\mathrm{FH}}' = \bigg{(}e^{-J\sum_i n_{i, \uparrow} n_{i, \downarrow} } \bigg{)} H_{\mathrm{FH}} \bigg{(} e^{J\sum_j n_{j, \uparrow} n_{j, \downarrow} } \bigg{)},
\end{equation}
where $J$ defines the strength of the transformation. This can be simplified to~\cite{dobrautz2019TC_hubbard}
\begin{equation}\label{Eq:TC_hubbard_3}
\begin{aligned}
    H_{\mathrm{FH}}' &= H_{\mathrm{FH}} - t \sum_{\langle i, j\rangle, \sigma} \bigg{(} a_{i,\sigma}^\dag a_{j, \sigma} \times \\
    [&(e^J - 1)n_{j, \bar{\sigma}}
    + (e^{-J} - 1) n_{i, \bar{\sigma}} 
    -2(\mathrm{cosh}(J)-1)n_{i, \bar{\sigma}} n_{j, \bar{\sigma}} ] \bigg{)} 
\end{aligned}
\end{equation}
where $\bar{\sigma}$ denotes the spin opposite to $\sigma$. 

Numerical simulations were carried out on $2 \times 2$ and $3 \times 2$ lattices, which can be mapped onto 8 and 12 qubits (respectively). The occupancy of each spin-lattice site is stored by the $\ket{0}$ or $\ket{1}$ value of a corresponding qubit. This is known as the Jordan-Wigner transformation, which can be used to map the fermionic operators to qubit operators
\begin{equation}
	\begin{aligned}
		a_p &= \frac{1}{2}(X_p + iY_p) \otimes Z_{p-1}\otimes \dots\otimes  Z_{0},\\
		a_p^\dag &= \frac{1}{2}(X_p - iY_p) \otimes Z_{p-1}\otimes \dots\otimes  Z_{0}.
	\end{aligned}
\end{equation}
The Hamiltonians can be written as a linear combination of tensor products of Pauli operators
\begin{equation}
	H = \sum_i \alpha_i P_i = \sum_i \alpha_i \prod_j \sigma_j^i,
\end{equation}
where $\alpha_i$ are scalar coefficients, $\sigma_j^i$ represents the Pauli operator in term $i$ acting on qubit $j$. The unmodified and TC Hamiltonians were generated and mapped to qubit operators using OpenFermion, a Python library for constructing chemistry Hamiltonians for use in quantum simulations~\cite{mcclean2017openfermion}.

The ansatz circuit used is based on the Hamiltonian variational ansatz~\cite{PhysRevA.92.042303}, which has proven effective in previous numerical simulations of applying the VQE to the FH model~\cite{cai2019hubbard, reiner2018hubbardground,cade2019fermihubbard}. We first prepare the quantum register in an eigenstate of the non-interacting ($U=0$) unmodified Hamiltonian. We choose the lowest energy eigenstate with the same particle number as the ground state of the full Hamiltonian. This state can easily be constructed using a network of Givens rotations~\cite{wecker2015hubbard, KivLinearDepth, jiang2018hubbard}. The ansatz is then constructed from repeated layers of a Trotterised decomposition of the time evolution operator $e^{-iHt}$. We assign a different parameter to each Pauli rotation in the Trotterised decomposition. The ansatz is given by
\begin{equation}
    U = \prod_l^L \prod_j e^{i \theta_j^l P_j}
\end{equation}
where $L$ denotes the number of layers in the ansatz, and $P_j$ are the Pauli strings in the Hamiltonian. Each of these Pauli rotations can be decomposed into CNOT gates and single qubit rotations~\cite{nielsen2002quantum}. We include an additional parameter to represent the global phase of the wavefunction. The importance of tracking the global phase during variational imaginary time evolution was discussed by \textcite{yuan2018variationaltheory}. The initial values of the parameters are perturbed from zero by a random perturbation upper bounded by $0.02\pi$, in order to prevent the method from becoming trapped in local minima around the non-interacting initial state. \\

\begin{figure}[h]\centering
\includegraphics[width=1.1\columnwidth]{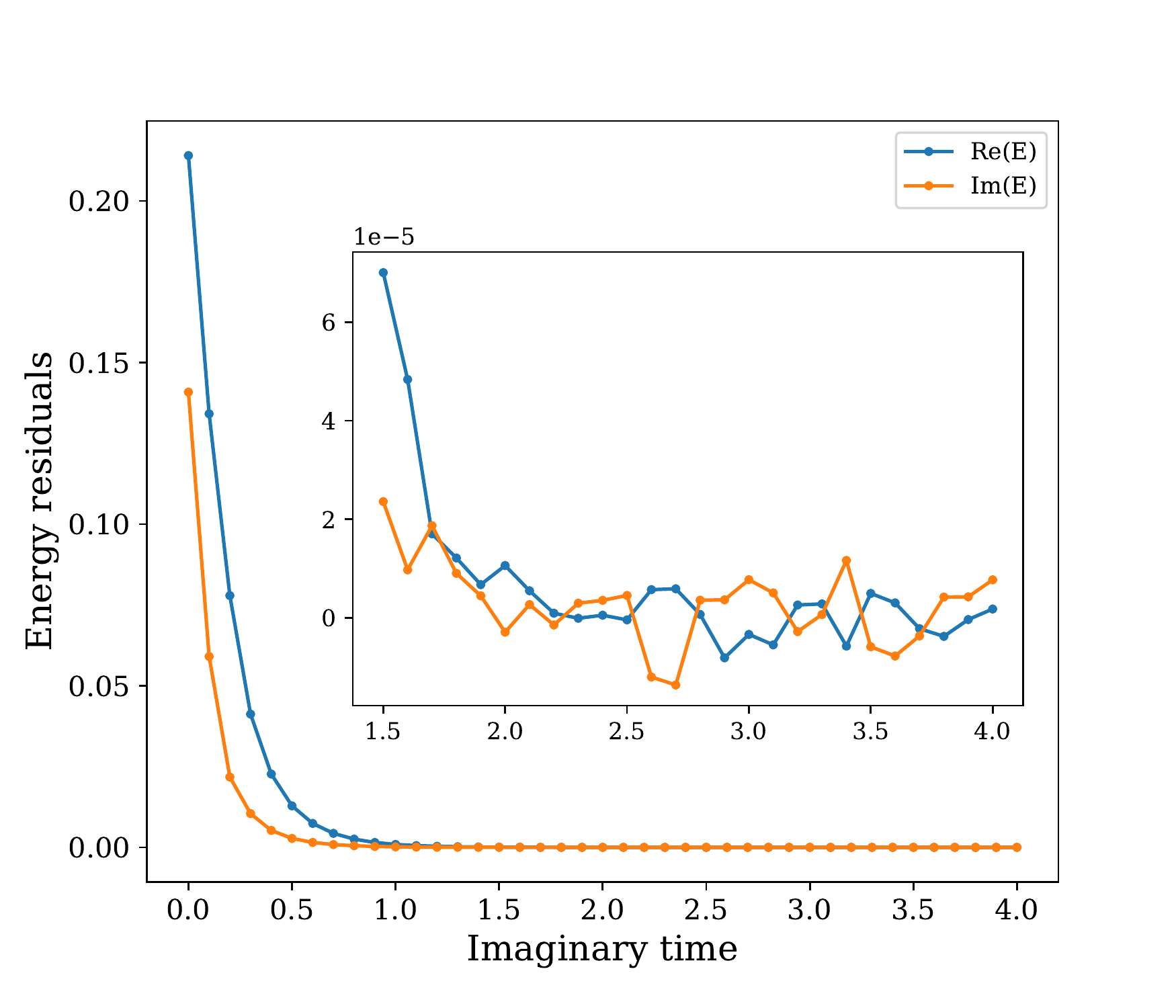}
 \caption{A random instance of finding the ground state of the TC Hamiltonian (with $J=-0.5$) for a $2 \times 2$ Fermi-Hubbard model, with imaginary time evolution. Energies were recorded every $10$ timesteps of the algorithm. An ansatz circuit with 3 layers was used. The inset plot shows the evolution of the energy during the final $62.5~\%$ of the runtime.} \label{Fig:Timerun_8}
\end{figure}

As a first step, we verify that the algorithm is capable of finding the ground state of the TC Hamiltonian. For this initial trial calculation, we target the ground state of the $2 \times 2$ FH model. We set $J=-0.5$, similar to the values used in Refs.~\cite{tsuneyuki2008TChubbard,dobrautz2019TC_hubbard}, as it is not necessary to use the optimal $J$ value for this calculation. In Fig.~\ref{Fig:Timerun_8}, we show a randomly chosen run of the imaginary time algorithm, applied to the $2 \times 2$ TC Hamiltonian, with a 3-layer ansatz. We observe that the energy rapidly converges towards the ground state value, and that the imaginary part of the energy decays exponentially as the method moves forwards in imaginary time. The inset of Fig.~\ref{Fig:Timerun_8} shows that although the algorithm is no longer strictly variational (the real part of the measured energy no longer upper bounds the ground state value), the real and imaginary parts of the energy converge towards their true values. 

\begin{figure}[h]\centering
\includegraphics[width=1.0\columnwidth]{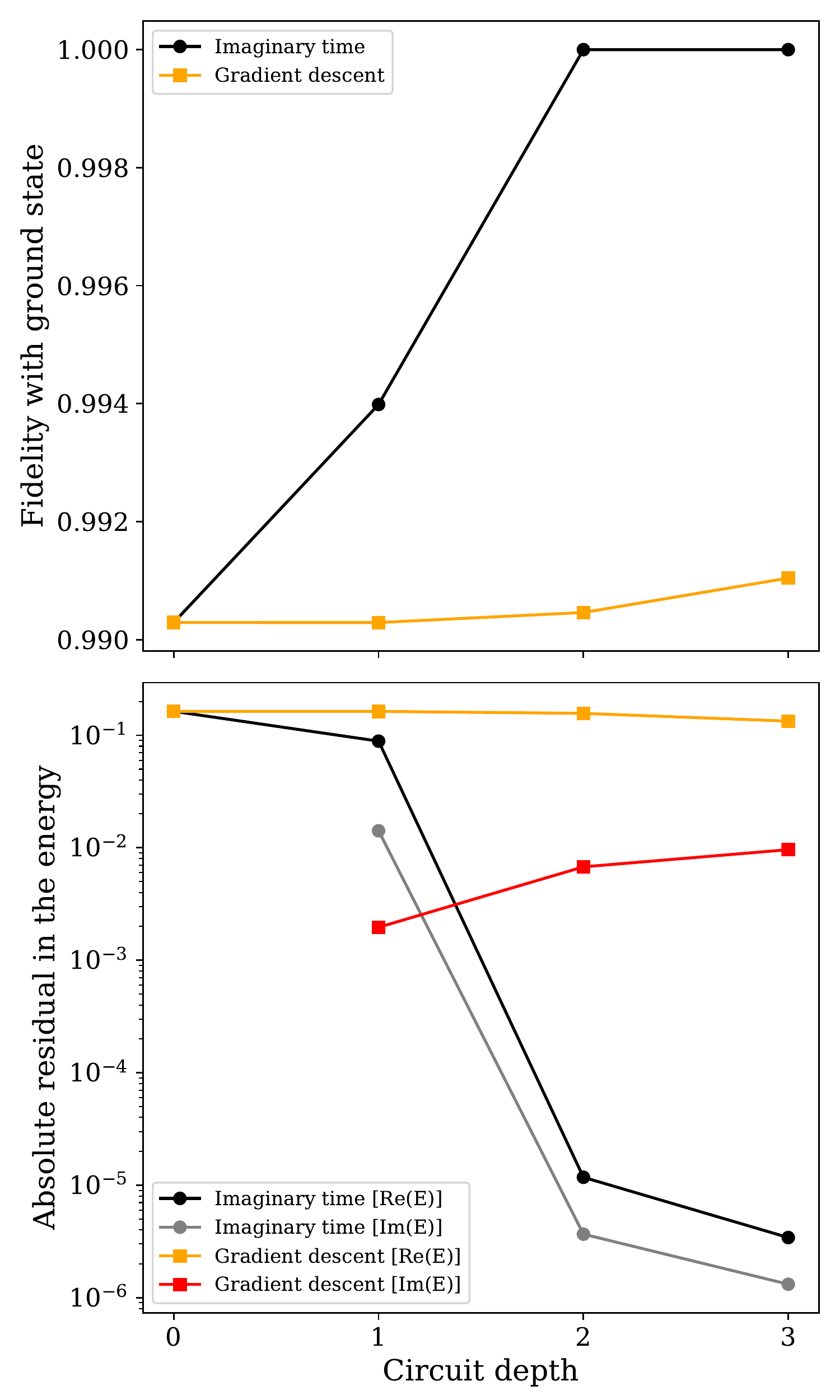}
 \caption{Finding the ground state of the TC Hamiltonian (with $J=-0.5$) for a $2 \times 2$ Fermi-Hubbard model. Lines are included to guide the eye. The upper plot shows fidelity between the TC right-hand ground state, and the states obtained from imaginary time evolution, or gradient descent, for a given number of ansatz layers. A circuit depth of zero denotes the non-interacting initial state. The lower plot shows the absolute residuals in the real and imaginary parts of the ground state energy value obtained from imaginary time evolution, and gradient descent. The non-interacting initial state gives a TC energy with a negligible imaginary part; this is not plotted, as it would distort the scale.} \label{Fig:Fids_8}
\end{figure}

We present the average results of ten repetitions of such calculations in Fig.~\ref{Fig:Fids_8}, for a range of ansatz depths. The upper plot of Fig.~\ref{Fig:Fids_8} shows that as the circuit depth is increased, the imaginary time algorithm is able to find better approximations to the right-hand ground state of the TC Hamiltonian. In contrast, we see that the fidelity obtained using gradient descent does not improve as the circuit depth increases, suggesting that algorithms based solely on cost function minimisation will struggle to find the ground state of the TC Hamiltonian. The number of parameters used is $28L +1$, where $L$ is the number of ansatz layers. This is much smaller than the Hilbert space dimension of $2^8 = 256$, for all circuit depths tested. This suggests that our method can still be effective, even if the ansatz used is unable to generate all possible states in the Hilbert space. The lower plot in Fig.~\ref{Fig:Fids_8} shows the (absolute) real and imaginary components of the energy residual. Once again, we see that while imaginary time evolution is able to find better approximations of the ground state as the circuit depth is increased, gradient descent barely improves upon the non-interacting initial state. It is interesting to compare the gradient descent and imaginary time datapoints for a single layer ansatz. Although the methods achieve similar energy values, we observe that imaginary time evolution achieves a much higher fidelity with the true TC ground state. This emphasises how our method attempts to evolve the state in imaginary time, rather than simply optimising the energy. \\

\begin{figure}[h]\centering
\includegraphics[width=1.1\columnwidth]{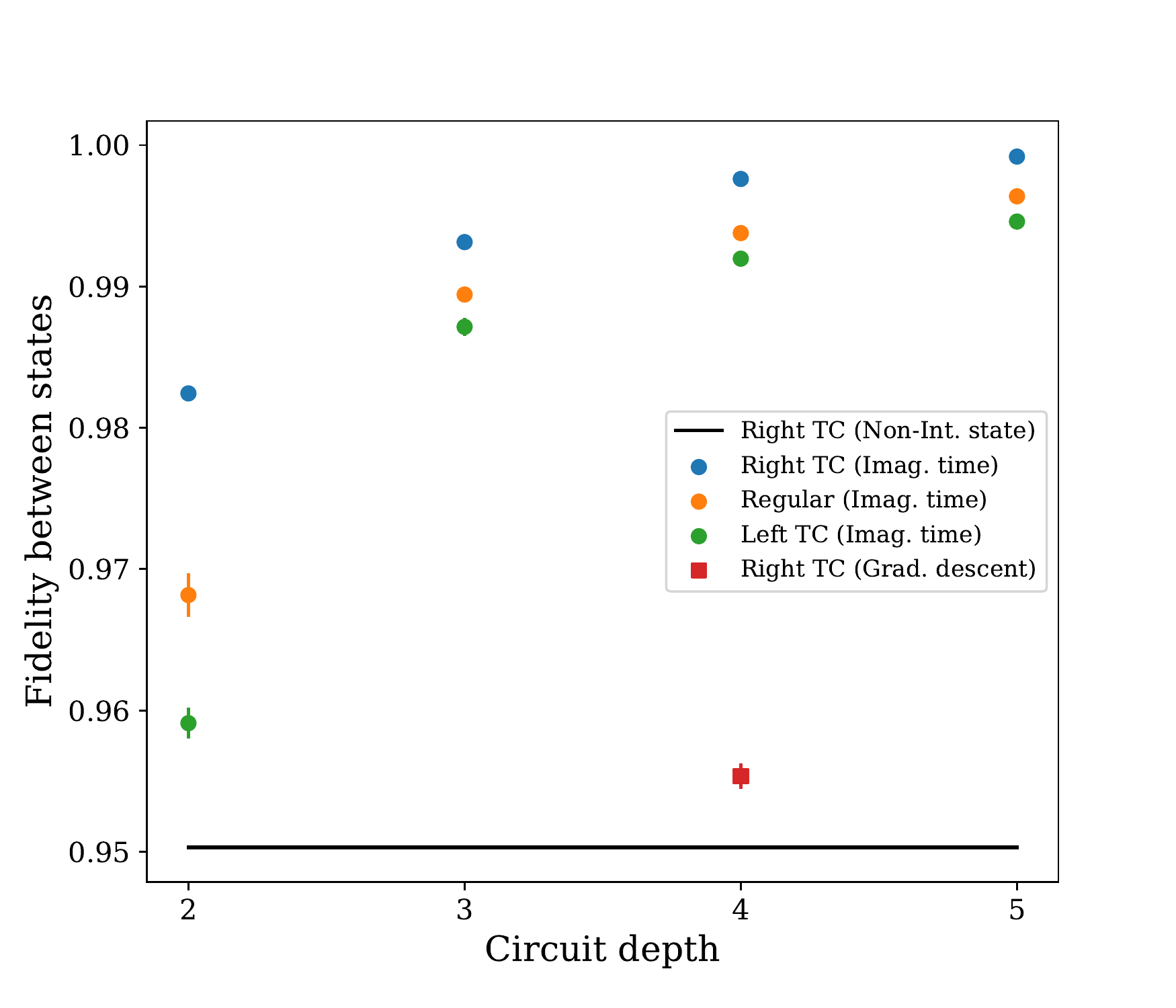}
 \caption{Finding the ground states of the regular and TC Hamiltonians for a $3 \times 2$ Fermi-Hubbard model. The TC Hamiltonian has $J=-0.6$. The figure shows fidelities between the target states specified in the legend, and the states generated by the method given in brackets in the legend. `Right (left) TC' denotes the right (left)-hand lowest energy eigenvector of the TC Hamiltonian. `Regular' denotes the lowest energy eigenvector of the unmodified Hamiltonian. Fidelities were obtained from 10 repetitions of each method (standard error bars are present, but small).} \label{Fig:Fids_12}
\end{figure}

We also consider the $3 \times 2$ FH model, as a more thorough test of using the imaginary time algorithm to find the ground states of the TC Hamiltonian. We first optimise the $J$ value of the TC Hamiltonian. We choose the value which gives the highest fidelity to the right-hand TC eigenvector when an ansatz circuit with two layers (the number of parameters is given by $46L +1$) is used. The optimal $J$ value was found to be $J=-0.6$. Although optimising the $J$ value in this way is not efficient, or tractable, for larger system sizes, we note that more efficient classical methods to optimise $J$ have been developed~\cite{tsuneyuki2008TChubbard, dobrautz2019TC_hubbard}. Moreover, the expansion of the ground state as a Slater determinant wavefunction can be used to compensate for any shortcomings induced by choosing a non-optimal $J$ value. Fig.~\ref{Fig:Fids_12} shows the fidelities obtained by targeting the right-hand and left-hand eigenvectors of the TC Hamiltonian, and the ground state of the regular Hamiltonian. We can see that the non-interacting initial state (before random perturbations are applied) already has a large overlap with the TC right-hand ground state. This fidelity is much higher than the fidelity of 0.81 between the non-interacting initial state and the ground state of the unmodified Hamiltonian. This highlights the additional correlation accounted for when using the TC Hamiltonian.

We see from Fig.~\ref{Fig:Fids_12} that once again, gradient descent is unable to find the right-hand ground state of the TC Hamiltonian. In contrast, imaginary time evolution is able to successfully find both the right-hand and left-hand eigenstates of the TC Hamiltonian. We can see that for a given circuit depth, we obtain the highest fidelity by applying imaginary time evolution to the TC Hamiltonian to find the TC right-hand eigenvector. A lower fidelity is obtained when using imaginary time evolution to find the ground state of the unmodified Hamiltonian. We obtain even lower fidelities when using imaginary time evolution to find the TC left-hand eigenvector. We attribute these effects to the changes that the TC transformation makes to the structure of the right-hand and left-hand eigenvectors, as discussed in Sec.~\ref{Sec:TC_approach}. As noted by \textcite{dobrautz2019TC_hubbard}, the TC transformation makes the right-hand eigenvector more compact, while making the left-hand eigenvector less compact. Their definition of compactness is closely related to the fidelity between a state composed of a mean-field solution plus double fermionic excitations, and the true ground state. Their results essentially show that fewer excitations are required to accurately approximate the right-hand TC eigenstate than the unmodified Hamiltonian's groundstate, which in turn, requires fewer excitations to approximate than the left-hand eigenstate. This is reflected in our results, where the number of ansatz layers can be likened to the number of excitations considered. As such, using the TC Hamiltonian may enable us to obtain accurate energies from a quantum simulation, using a lower depth circuit. As quantum hardware is currently limited by the number of gates that can be applied before noise corrupts the results of the calculation, this highlights the value of the TC method. However, while the TC Hamiltonian improves our ability to find the right-hand eigenstate, it may hamper our ability to find the left-hand eigenstate. While this is not problematic when measuring only the energy, it introduces difficulties when measuring other observables, as discussed in Sec.~\ref{Sec:TC_approach}.

\section{Discussion}\label{Sec:Discussion}
In this work, we have investigated introducing the transcorrelated (TC) method to quantum algorithms for solving the electronic structure problem. As discussed in Sec.~\ref{Sec:TC_approach}, the TC method transfers dynamic correlation from the right-hand eigenstates, to the Hamiltonian, through a similarity transform. This enables accurate energies to be obtained using smaller single particle basis sets and/or more compact Slater determinant expansions. As the TC Hamiltonian is no longer Hermitian, it lacks a variational lower bound, and so appears difficult to integrate with existing quantum algorithms, such as the VQE. Motivated by the success of imaginary time-like methods, such as FCIQMC, we investigated solving the TC Hamiltonian using a quantum algorithm that implements ansatz-based imaginary time evolution. We performed numerical simulations of this method applied to small Fermi-Hubbard models, verifying its compatibility with the TC method. We observe that the use of a TC Hamiltonian can reduce the circuit depth at which we are able to find high accuracy approximations of the ground state eigenvalue, in agreement with previous (classical) investigations of the TC method~\cite{dobrautz2019TC_hubbard}. As quantum simulations are currently depth-limited due to noise, this may enable more accurate results to be obtained from an imperfect quantum computer. We note that the TC method is more beneficial for molecular systems than for the Fermi-Hubbard model (as the TC transformation on the former is performed before projection onto a basis set, as described in Sec.~\ref{Sec:TC_approach}). As such, our method may be even more impactful on quantum simulations of molecular systems, where it may reduce the number of qubits required for the simulation, as well as the gate depth needed to generate an accurate estimate of the ground state. \\

There are a number of open questions surrounding the use of the TC method with quantum imaginary time evolution. We note that the TC method is an active area of research in classical computational chemistry, with a number of challenges to overcome. Most notably, the TC Hamiltonian for molecules in a Gaussian basis contains up to $O(M^6)$ terms, which makes evaluating the basis function integrals the current bottleneck for the method~\cite{cohen2019TCatoms}. As discussed earlier, while the TC method makes the right-hand eigenvectors easier to obtain, the left-hand eigenvectors become more difficult to construct. It is not yet known how to best overcome this problem, when measuring observables other than the energy. Similarly, as the right-hand eigenstates of the TC Hamiltonian are no longer orthogonal to each other, quantum algorithms for finding excited states based on orthogonality~\cite{higgott2018variational,jones2019ImagExcited} are no longer applicable. While other quantum algorithms to find excited states, such as the quantum subspace expansion~\cite{PhysRevA.95.042308,PhysRevX.8.011021}, should still be compatible with the TC method, it will be interesting to determine what other limitations the TC Hamiltonian introduces. It would also be valuable to optimise the approach taken thus far. For example, developing improved ansatz circuits, or investigating alternative TC transformations~\cite{neuscamman2011TC_transform,Wahlen2015TC_transforms}, which may be more effective than the methods tested in this work. 

The promising results obtained from this initial study suggest that the transcorrelated method could benefit near-term quantum simulations, providing improved accuracy when estimating ground state energies, at a reduced cost.\\

\noindent \textbf{Acknowledgements\\}
We thank Simon Benjamin for helpful discussions on this work. This work was supported by the EPSRC National Quantum Technology
Hub in Networked Quantum Information Technology
(EP/M013243/1) and the QCS Hub (EP/T001062/1). The authors would like to acknowledge the use of the University of Oxford Advanced Research Computing (ARC) facility in carrying out this work. http://dx.doi.org/10.5281/zenodo.22558 .\\

\bibliography{ChemReviewBib}\newpage

\begin{thebibliography}{78}%
\makeatletter
\providecommand \@ifxundefined [1]{%
 \@ifx{#1\undefined}
}%
\providecommand \@ifnum [1]{%
 \ifnum #1\expandafter \@firstoftwo
 \else \expandafter \@secondoftwo
 \fi
}%
\providecommand \@ifx [1]{%
 \ifx #1\expandafter \@firstoftwo
 \else \expandafter \@secondoftwo
 \fi
}%
\providecommand \natexlab [1]{#1}%
\providecommand \enquote  [1]{``#1''}%
\providecommand \bibnamefont  [1]{#1}%
\providecommand \bibfnamefont [1]{#1}%
\providecommand \citenamefont [1]{#1}%
\providecommand \href@noop [0]{\@secondoftwo}%
\providecommand \href [0]{\begingroup \@sanitize@url \@href}%
\providecommand \@href[1]{\@@startlink{#1}\@@href}%
\providecommand \@@href[1]{\endgroup#1\@@endlink}%
\providecommand \@sanitize@url [0]{\catcode `\\12\catcode `\$12\catcode
  `\&12\catcode `\#12\catcode `\^12\catcode `\_12\catcode `\%12\relax}%
\providecommand \@@startlink[1]{}%
\providecommand \@@endlink[0]{}%
\providecommand \url  [0]{\begingroup\@sanitize@url \@url }%
\providecommand \@url [1]{\endgroup\@href {#1}{\urlprefix }}%
\providecommand \urlprefix  [0]{URL }%
\providecommand \Eprint [0]{\href }%
\providecommand \doibase [0]{http://dx.doi.org/}%
\providecommand \selectlanguage [0]{\@gobble}%
\providecommand \bibinfo  [0]{\@secondoftwo}%
\providecommand \bibfield  [0]{\@secondoftwo}%
\providecommand \translation [1]{[#1]}%
\providecommand \BibitemOpen [0]{}%
\providecommand \bibitemStop [0]{}%
\providecommand \bibitemNoStop [0]{.\EOS\space}%
\providecommand \EOS [0]{\spacefactor3000\relax}%
\providecommand \BibitemShut  [1]{\csname bibitem#1\endcsname}%
\let\auto@bib@innerbib\@empty
\bibitem [{\citenamefont {Feynman}(1982)}]{feynman1982simulating}%
  \BibitemOpen
  \bibfield  {author} {\bibinfo {author} {\bibfnamefont {R.~P.}\ \bibnamefont
  {Feynman}},\ }\href@noop {} {\bibfield  {journal} {\bibinfo  {journal}
  {International journal of theoretical physics}\ }\textbf {\bibinfo {volume}
  {21}},\ \bibinfo {pages} {467} (\bibinfo {year} {1982})}\BibitemShut
  {NoStop}%
\bibitem [{\citenamefont {Cao}\ \emph {et~al.}(2019)\citenamefont {Cao},
  \citenamefont {Romero}, \citenamefont {Olson}, \citenamefont {Degroote},
  \citenamefont {Johnson}, \citenamefont {Kieferová}, \citenamefont
  {Kivlichan}, \citenamefont {Menke}, \citenamefont {Peropadre}, \citenamefont
  {Sawaya}, \citenamefont {Sim}, \citenamefont {Veis},\ and\ \citenamefont
  {Aspuru-Guzik}}]{cao2018chemreview}%
  \BibitemOpen
  \bibfield  {author} {\bibinfo {author} {\bibfnamefont {Y.}~\bibnamefont
  {Cao}}, \bibinfo {author} {\bibfnamefont {J.}~\bibnamefont {Romero}},
  \bibinfo {author} {\bibfnamefont {J.~P.}\ \bibnamefont {Olson}}, \bibinfo
  {author} {\bibfnamefont {M.}~\bibnamefont {Degroote}}, \bibinfo {author}
  {\bibfnamefont {P.~D.}\ \bibnamefont {Johnson}}, \bibinfo {author}
  {\bibfnamefont {M.}~\bibnamefont {Kieferová}}, \bibinfo {author}
  {\bibfnamefont {I.~D.}\ \bibnamefont {Kivlichan}}, \bibinfo {author}
  {\bibfnamefont {T.}~\bibnamefont {Menke}}, \bibinfo {author} {\bibfnamefont
  {B.}~\bibnamefont {Peropadre}}, \bibinfo {author} {\bibfnamefont {N.~P.~D.}\
  \bibnamefont {Sawaya}}, \bibinfo {author} {\bibfnamefont {S.}~\bibnamefont
  {Sim}}, \bibinfo {author} {\bibfnamefont {L.}~\bibnamefont {Veis}}, \ and\
  \bibinfo {author} {\bibfnamefont {A.}~\bibnamefont {Aspuru-Guzik}},\ }\href
  {\doibase 10.1021/acs.chemrev.8b00803} {\bibfield  {journal} {\bibinfo
  {journal} {Chemical Reviews}\ }\textbf {\bibinfo {volume} {119}},\ \bibinfo
  {pages} {10856} (\bibinfo {year} {2019})},\ \bibinfo {note} {pMID:
  31469277},\ \Eprint
  {http://arxiv.org/abs/https://doi.org/10.1021/acs.chemrev.8b00803}
  {https://doi.org/10.1021/acs.chemrev.8b00803} \BibitemShut {NoStop}%
\bibitem [{\citenamefont {McArdle}\ \emph {et~al.}(2020)\citenamefont
  {McArdle}, \citenamefont {Endo}, \citenamefont {Aspuru-Guzik}, \citenamefont
  {Benjamin},\ and\ \citenamefont {Yuan}}]{mcardle2020review}%
  \BibitemOpen
  \bibfield  {author} {\bibinfo {author} {\bibfnamefont {S.}~\bibnamefont
  {McArdle}}, \bibinfo {author} {\bibfnamefont {S.}~\bibnamefont {Endo}},
  \bibinfo {author} {\bibfnamefont {A.}~\bibnamefont {Aspuru-Guzik}}, \bibinfo
  {author} {\bibfnamefont {S.~C.}\ \bibnamefont {Benjamin}}, \ and\ \bibinfo
  {author} {\bibfnamefont {X.}~\bibnamefont {Yuan}},\ }\href {\doibase
  10.1103/RevModPhys.92.015003} {\bibfield  {journal} {\bibinfo  {journal}
  {Rev. Mod. Phys.}\ }\textbf {\bibinfo {volume} {92}},\ \bibinfo {pages}
  {015003} (\bibinfo {year} {2020})}\BibitemShut {NoStop}%
\bibitem [{\citenamefont {Bauer}\ \emph {et~al.}(2020)\citenamefont {Bauer},
  \citenamefont {Bravyi}, \citenamefont {Motta},\ and\ \citenamefont
  {Chan}}]{bauer2020QchemReview}%
  \BibitemOpen
  \bibfield  {author} {\bibinfo {author} {\bibfnamefont {B.}~\bibnamefont
  {Bauer}}, \bibinfo {author} {\bibfnamefont {S.}~\bibnamefont {Bravyi}},
  \bibinfo {author} {\bibfnamefont {M.}~\bibnamefont {Motta}}, \ and\ \bibinfo
  {author} {\bibfnamefont {G.~K.-L.}\ \bibnamefont {Chan}},\ }\href@noop {} {\
  (\bibinfo {year} {2020})},\ \Eprint {http://arxiv.org/abs/arXiv:2001.03685}
  {arXiv:2001.03685} \BibitemShut {NoStop}%
\bibitem [{\citenamefont {Gr\"uneis}\ \emph {et~al.}(2017)\citenamefont
  {Gr\"uneis}, \citenamefont {Hirata}, \citenamefont {Ohnishi},\ and\
  \citenamefont {Ten-no}}]{gruneis2017correlatedperspective}%
  \BibitemOpen
  \bibfield  {author} {\bibinfo {author} {\bibfnamefont {A.}~\bibnamefont
  {Gr\"uneis}}, \bibinfo {author} {\bibfnamefont {S.}~\bibnamefont {Hirata}},
  \bibinfo {author} {\bibfnamefont {Y.-y.}\ \bibnamefont {Ohnishi}}, \ and\
  \bibinfo {author} {\bibfnamefont {S.}~\bibnamefont {Ten-no}},\ }\href
  {\doibase 10.1063/1.4976974} {\bibfield  {journal} {\bibinfo  {journal} {The
  Journal of Chemical Physics}\ }\textbf {\bibinfo {volume} {146}},\ \bibinfo
  {pages} {080901} (\bibinfo {year} {2017})},\ \Eprint
  {http://arxiv.org/abs/https://doi.org/10.1063/1.4976974}
  {https://doi.org/10.1063/1.4976974} \BibitemShut {NoStop}%
\bibitem [{\citenamefont {Hattig}\ \emph {et~al.}(2012)\citenamefont {Hattig},
  \citenamefont {Klopper}, \citenamefont {Köhn},\ and\ \citenamefont
  {Tew}}]{hattig2012correlatedreview}%
  \BibitemOpen
  \bibfield  {author} {\bibinfo {author} {\bibfnamefont {C.}~\bibnamefont
  {Hattig}}, \bibinfo {author} {\bibfnamefont {W.}~\bibnamefont {Klopper}},
  \bibinfo {author} {\bibfnamefont {A.}~\bibnamefont {Köhn}}, \ and\ \bibinfo
  {author} {\bibfnamefont {D.~P.}\ \bibnamefont {Tew}},\ }\href {\doibase
  10.1021/cr200168z} {\bibfield  {journal} {\bibinfo  {journal} {Chemical
  Reviews}\ }\textbf {\bibinfo {volume} {112}},\ \bibinfo {pages} {4} (\bibinfo
  {year} {2012})},\ \bibinfo {note} {pMID: 22206503},\ \Eprint
  {http://arxiv.org/abs/https://doi.org/10.1021/cr200168z}
  {https://doi.org/10.1021/cr200168z} \BibitemShut {NoStop}%
\bibitem [{\citenamefont {Kong}\ \emph {et~al.}(2012)\citenamefont {Kong},
  \citenamefont {Bischoff},\ and\ \citenamefont
  {Valeev}}]{Kong2012explicitlycorrelated}%
  \BibitemOpen
  \bibfield  {author} {\bibinfo {author} {\bibfnamefont {L.}~\bibnamefont
  {Kong}}, \bibinfo {author} {\bibfnamefont {F.~A.}\ \bibnamefont {Bischoff}},
  \ and\ \bibinfo {author} {\bibfnamefont {E.~F.}\ \bibnamefont {Valeev}},\
  }\href {\doibase 10.1021/cr200204r} {\bibfield  {journal} {\bibinfo
  {journal} {Chemical Reviews}\ }\textbf {\bibinfo {volume} {112}},\ \bibinfo
  {pages} {75} (\bibinfo {year} {2012})}\BibitemShut {NoStop}%
\bibitem [{\citenamefont {Boys}\ \emph
  {et~al.}(1969{\natexlab{a}})\citenamefont {Boys}, \citenamefont {Handy},\
  and\ \citenamefont {Linnett}}]{boys1969tc1}%
  \BibitemOpen
  \bibfield  {author} {\bibinfo {author} {\bibfnamefont {S.~F.}\ \bibnamefont
  {Boys}}, \bibinfo {author} {\bibfnamefont {N.~C.}\ \bibnamefont {Handy}}, \
  and\ \bibinfo {author} {\bibfnamefont {J.~W.}\ \bibnamefont {Linnett}},\
  }\href {\doibase 10.1098/rspa.1969.0038} {\bibfield  {journal} {\bibinfo
  {journal} {Proceedings of the Royal Society of London. A. Mathematical and
  Physical Sciences}\ }\textbf {\bibinfo {volume} {309}},\ \bibinfo {pages}
  {209} (\bibinfo {year} {1969}{\natexlab{a}})},\ \Eprint
  {http://arxiv.org/abs/https://royalsocietypublishing.org/doi/pdf/10.1098/rspa.1969.0038}
  {https://royalsocietypublishing.org/doi/pdf/10.1098/rspa.1969.0038}
  \BibitemShut {NoStop}%
\bibitem [{\citenamefont {Boys}\ \emph
  {et~al.}(1969{\natexlab{b}})\citenamefont {Boys}, \citenamefont {Handy},\
  and\ \citenamefont {Linnett}}]{boys1969tc2}%
  \BibitemOpen
  \bibfield  {author} {\bibinfo {author} {\bibfnamefont {S.~F.}\ \bibnamefont
  {Boys}}, \bibinfo {author} {\bibfnamefont {N.~C.}\ \bibnamefont {Handy}}, \
  and\ \bibinfo {author} {\bibfnamefont {J.~W.}\ \bibnamefont {Linnett}},\
  }\href {\doibase 10.1098/rspa.1969.0061} {\bibfield  {journal} {\bibinfo
  {journal} {Proceedings of the Royal Society of London. A. Mathematical and
  Physical Sciences}\ }\textbf {\bibinfo {volume} {310}},\ \bibinfo {pages}
  {43} (\bibinfo {year} {1969}{\natexlab{b}})},\ \Eprint
  {http://arxiv.org/abs/https://royalsocietypublishing.org/doi/pdf/10.1098/rspa.1969.0061}
  {https://royalsocietypublishing.org/doi/pdf/10.1098/rspa.1969.0061}
  \BibitemShut {NoStop}%
\bibitem [{\citenamefont {McArdle}\ \emph {et~al.}(2019)\citenamefont
  {McArdle}, \citenamefont {Jones}, \citenamefont {Endo}, \citenamefont {Li},
  \citenamefont {Benjamin},\ and\ \citenamefont
  {Yuan}}]{mcardle2018variational}%
  \BibitemOpen
  \bibfield  {author} {\bibinfo {author} {\bibfnamefont {S.}~\bibnamefont
  {McArdle}}, \bibinfo {author} {\bibfnamefont {T.}~\bibnamefont {Jones}},
  \bibinfo {author} {\bibfnamefont {S.}~\bibnamefont {Endo}}, \bibinfo {author}
  {\bibfnamefont {Y.}~\bibnamefont {Li}}, \bibinfo {author} {\bibfnamefont
  {S.~C.}\ \bibnamefont {Benjamin}}, \ and\ \bibinfo {author} {\bibfnamefont
  {X.}~\bibnamefont {Yuan}},\ }\href {\doibase 10.1038/s41534-019-0187-2}
  {\bibfield  {journal} {\bibinfo  {journal} {npj Quantum Information}\
  }\textbf {\bibinfo {volume} {5}} (\bibinfo {year} {2019}),\
  10.1038/s41534-019-0187-2}\BibitemShut {NoStop}%
\bibitem [{\citenamefont {Luo}\ and\ \citenamefont
  {Alavi}(2018)}]{luo2018TC_FCIQMC_planewave}%
  \BibitemOpen
  \bibfield  {author} {\bibinfo {author} {\bibfnamefont {H.}~\bibnamefont
  {Luo}}\ and\ \bibinfo {author} {\bibfnamefont {A.}~\bibnamefont {Alavi}},\
  }\href {\doibase 10.1021/acs.jctc.7b01257} {\bibfield  {journal} {\bibinfo
  {journal} {Journal of Chemical Theory and Computation}\ }\textbf {\bibinfo
  {volume} {14}},\ \bibinfo {pages} {1403} (\bibinfo {year} {2018})},\ \bibinfo
  {note} {pMID: 29431996},\ \Eprint
  {http://arxiv.org/abs/https://doi.org/10.1021/acs.jctc.7b01257}
  {https://doi.org/10.1021/acs.jctc.7b01257} \BibitemShut {NoStop}%
\bibitem [{\citenamefont {Dobrautz}\ \emph {et~al.}(2019)\citenamefont
  {Dobrautz}, \citenamefont {Luo},\ and\ \citenamefont
  {Alavi}}]{dobrautz2019TC_hubbard}%
  \BibitemOpen
  \bibfield  {author} {\bibinfo {author} {\bibfnamefont {W.}~\bibnamefont
  {Dobrautz}}, \bibinfo {author} {\bibfnamefont {H.}~\bibnamefont {Luo}}, \
  and\ \bibinfo {author} {\bibfnamefont {A.}~\bibnamefont {Alavi}},\ }\href
  {\doibase 10.1103/PhysRevB.99.075119} {\bibfield  {journal} {\bibinfo
  {journal} {Phys. Rev. B}\ }\textbf {\bibinfo {volume} {99}},\ \bibinfo
  {pages} {075119} (\bibinfo {year} {2019})}\BibitemShut {NoStop}%
\bibitem [{\citenamefont {Cohen}\ \emph {et~al.}(2019)\citenamefont {Cohen},
  \citenamefont {Luo}, \citenamefont {Guther}, \citenamefont {Dobrautz},
  \citenamefont {Tew},\ and\ \citenamefont {Alavi}}]{cohen2019TCatoms}%
  \BibitemOpen
  \bibfield  {author} {\bibinfo {author} {\bibfnamefont {A.~J.}\ \bibnamefont
  {Cohen}}, \bibinfo {author} {\bibfnamefont {H.}~\bibnamefont {Luo}}, \bibinfo
  {author} {\bibfnamefont {K.}~\bibnamefont {Guther}}, \bibinfo {author}
  {\bibfnamefont {W.}~\bibnamefont {Dobrautz}}, \bibinfo {author}
  {\bibfnamefont {D.~P.}\ \bibnamefont {Tew}}, \ and\ \bibinfo {author}
  {\bibfnamefont {A.}~\bibnamefont {Alavi}},\ }\href {\doibase
  10.1063/1.5116024} {\bibfield  {journal} {\bibinfo  {journal} {The Journal of
  Chemical Physics}\ }\textbf {\bibinfo {volume} {151}},\ \bibinfo {pages}
  {061101} (\bibinfo {year} {2019})},\ \Eprint
  {http://arxiv.org/abs/https://doi.org/10.1063/1.5116024}
  {https://doi.org/10.1063/1.5116024} \BibitemShut {NoStop}%
\bibitem [{\citenamefont {Motta}\ \emph {et~al.}(2020)\citenamefont {Motta},
  \citenamefont {Gujarati}, \citenamefont {Rice}, \citenamefont {Kumar},
  \citenamefont {Masteran}, \citenamefont {Latone}, \citenamefont {Lee},
  \citenamefont {Valeev},\ and\ \citenamefont {Takeshita}}]{motta2020TC}%
  \BibitemOpen
  \bibfield  {author} {\bibinfo {author} {\bibfnamefont {M.}~\bibnamefont
  {Motta}}, \bibinfo {author} {\bibfnamefont {T.~P.}\ \bibnamefont {Gujarati}},
  \bibinfo {author} {\bibfnamefont {J.~E.}\ \bibnamefont {Rice}}, \bibinfo
  {author} {\bibfnamefont {A.}~\bibnamefont {Kumar}}, \bibinfo {author}
  {\bibfnamefont {C.}~\bibnamefont {Masteran}}, \bibinfo {author}
  {\bibfnamefont {J.~A.}\ \bibnamefont {Latone}}, \bibinfo {author}
  {\bibfnamefont {E.}~\bibnamefont {Lee}}, \bibinfo {author} {\bibfnamefont
  {E.~F.}\ \bibnamefont {Valeev}}, \ and\ \bibinfo {author} {\bibfnamefont
  {T.~Y.}\ \bibnamefont {Takeshita}},\ }\href@noop {} {\  (\bibinfo {year}
  {2020})},\ \Eprint {http://arxiv.org/abs/arXiv:2006.02488} {arXiv:2006.02488}
  \BibitemShut {NoStop}%
\bibitem [{\citenamefont {Yanai}\ and\ \citenamefont
  {Shiozaki}(2012)}]{yanai2012canonicalTC}%
  \BibitemOpen
  \bibfield  {author} {\bibinfo {author} {\bibfnamefont {T.}~\bibnamefont
  {Yanai}}\ and\ \bibinfo {author} {\bibfnamefont {T.}~\bibnamefont
  {Shiozaki}},\ }\href {\doibase 10.1063/1.3688225} {\bibfield  {journal}
  {\bibinfo  {journal} {The Journal of Chemical Physics}\ }\textbf {\bibinfo
  {volume} {136}},\ \bibinfo {pages} {084107} (\bibinfo {year} {2012})},\
  \Eprint {http://arxiv.org/abs/https://doi.org/10.1063/1.3688225}
  {https://doi.org/10.1063/1.3688225} \BibitemShut {NoStop}%
\bibitem [{\citenamefont {Helgaker}\ \emph {et~al.}(2014)\citenamefont
  {Helgaker}, \citenamefont {Jorgensen},\ and\ \citenamefont
  {Olsen}}]{helgaker2014molecular}%
  \BibitemOpen
  \bibfield  {author} {\bibinfo {author} {\bibfnamefont {T.}~\bibnamefont
  {Helgaker}}, \bibinfo {author} {\bibfnamefont {P.}~\bibnamefont {Jorgensen}},
  \ and\ \bibinfo {author} {\bibfnamefont {J.}~\bibnamefont {Olsen}},\
  }\href@noop {} {\emph {\bibinfo {title} {Molecular electronic-structure
  theory}}}\ (\bibinfo  {publisher} {John Wiley \& Sons},\ \bibinfo {year}
  {2014})\BibitemShut {NoStop}%
\bibitem [{\citenamefont {Szabo}\ and\ \citenamefont
  {Ostlund}(2012)}]{szabo2012modern}%
  \BibitemOpen
  \bibfield  {author} {\bibinfo {author} {\bibfnamefont {A.}~\bibnamefont
  {Szabo}}\ and\ \bibinfo {author} {\bibfnamefont {N.~S.}\ \bibnamefont
  {Ostlund}},\ }\href@noop {} {\emph {\bibinfo {title} {Modern quantum
  chemistry: introduction to advanced electronic structure theory}}}\ (\bibinfo
   {publisher} {Courier Corporation},\ \bibinfo {year} {2012})\BibitemShut
  {NoStop}%
\bibitem [{\citenamefont {Kato}(1957)}]{kato1957cusp}%
  \BibitemOpen
  \bibfield  {author} {\bibinfo {author} {\bibfnamefont {T.}~\bibnamefont
  {Kato}},\ }\href {\doibase 10.1002/cpa.3160100201} {\bibfield  {journal}
  {\bibinfo  {journal} {Communications on Pure and Applied Mathematics}\
  }\textbf {\bibinfo {volume} {10}},\ \bibinfo {pages} {151} (\bibinfo {year}
  {1957})},\ \Eprint
  {http://arxiv.org/abs/https://onlinelibrary.wiley.com/doi/pdf/10.1002/cpa.3160100201}
  {https://onlinelibrary.wiley.com/doi/pdf/10.1002/cpa.3160100201} \BibitemShut
  {NoStop}%
\bibitem [{\citenamefont {Nooijen}\ and\ \citenamefont
  {Bartlett}(1998)}]{nooijen1998InfinityElimination}%
  \BibitemOpen
  \bibfield  {author} {\bibinfo {author} {\bibfnamefont {M.}~\bibnamefont
  {Nooijen}}\ and\ \bibinfo {author} {\bibfnamefont {R.~J.}\ \bibnamefont
  {Bartlett}},\ }\href {\doibase 10.1063/1.477485} {\bibfield  {journal}
  {\bibinfo  {journal} {The Journal of Chemical Physics}\ }\textbf {\bibinfo
  {volume} {109}},\ \bibinfo {pages} {8232} (\bibinfo {year} {1998})},\ \Eprint
  {http://arxiv.org/abs/https://doi.org/10.1063/1.477485}
  {https://doi.org/10.1063/1.477485} \BibitemShut {NoStop}%
\bibitem [{\citenamefont {Hylleraas}(1929)}]{Hylleraas1929helium}%
  \BibitemOpen
  \bibfield  {author} {\bibinfo {author} {\bibfnamefont {E.~A.}\ \bibnamefont
  {Hylleraas}},\ }\href {\doibase 10.1007/BF01375457} {\bibfield  {journal}
  {\bibinfo  {journal} {Zeitschrift f{\"u}r Physik}\ }\textbf {\bibinfo
  {volume} {54}},\ \bibinfo {pages} {347} (\bibinfo {year} {1929})}\BibitemShut
  {NoStop}%
\bibitem [{\citenamefont {Slater}(1928)}]{slater1928rydberg}%
  \BibitemOpen
  \bibfield  {author} {\bibinfo {author} {\bibfnamefont {J.~C.}\ \bibnamefont
  {Slater}},\ }\href {\doibase 10.1103/PhysRev.31.333} {\bibfield  {journal}
  {\bibinfo  {journal} {Phys. Rev.}\ }\textbf {\bibinfo {volume} {31}},\
  \bibinfo {pages} {333} (\bibinfo {year} {1928})}\BibitemShut {NoStop}%
\bibitem [{\citenamefont {Kutzelnigg}(1985)}]{Kutzelnigg1985r12}%
  \BibitemOpen
  \bibfield  {author} {\bibinfo {author} {\bibfnamefont {W.}~\bibnamefont
  {Kutzelnigg}},\ }\href {\doibase 10.1007/BF00527669} {\bibfield  {journal}
  {\bibinfo  {journal} {Theoretica chimica acta}\ }\textbf {\bibinfo {volume}
  {68}},\ \bibinfo {pages} {445} (\bibinfo {year} {1985})}\BibitemShut
  {NoStop}%
\bibitem [{\citenamefont {Hirschfelder}(1963)}]{hirschfelder1963similarity}%
  \BibitemOpen
  \bibfield  {author} {\bibinfo {author} {\bibfnamefont {J.~O.}\ \bibnamefont
  {Hirschfelder}},\ }\href {\doibase 10.1063/1.1734157} {\bibfield  {journal}
  {\bibinfo  {journal} {The Journal of Chemical Physics}\ }\textbf {\bibinfo
  {volume} {39}},\ \bibinfo {pages} {3145} (\bibinfo {year}
  {1963})}\BibitemShut {NoStop}%
\bibitem [{\citenamefont {Tsuneyuki}(2008)}]{tsuneyuki2008TChubbard}%
  \BibitemOpen
  \bibfield  {author} {\bibinfo {author} {\bibfnamefont {S.}~\bibnamefont
  {Tsuneyuki}},\ }\href {\doibase 10.1143/PTPS.176.134} {\bibfield  {journal}
  {\bibinfo  {journal} {Progress of Theoretical Physics Supplement}\ }\textbf
  {\bibinfo {volume} {176}},\ \bibinfo {pages} {134} (\bibinfo {year}
  {2008})},\ \Eprint
  {http://arxiv.org/abs/https://academic.oup.com/ptps/article-pdf/doi/10.1143/PTPS.176.134/5321588/176-134.pdf}
  {https://academic.oup.com/ptps/article-pdf/doi/10.1143/PTPS.176.134/5321588/176-134.pdf}
  \BibitemShut {NoStop}%
\bibitem [{\citenamefont {Handy}(1971)}]{handy1971TranVariance}%
  \BibitemOpen
  \bibfield  {author} {\bibinfo {author} {\bibfnamefont {N.}~\bibnamefont
  {Handy}},\ }\href {\doibase 10.1080/00268977100101961} {\bibfield  {journal}
  {\bibinfo  {journal} {Molecular Physics}\ }\textbf {\bibinfo {volume} {21}},\
  \bibinfo {pages} {817} (\bibinfo {year} {1971})},\ \Eprint
  {http://arxiv.org/abs/https://doi.org/10.1080/00268977100101961}
  {https://doi.org/10.1080/00268977100101961} \BibitemShut {NoStop}%
\bibitem [{\citenamefont {Ten-no}(2000)}]{tenno2000transcorrelated}%
  \BibitemOpen
  \bibfield  {author} {\bibinfo {author} {\bibfnamefont {S.}~\bibnamefont
  {Ten-no}},\ }\href {\doibase https://doi.org/10.1016/S0009-2614(00)01066-6}
  {\bibfield  {journal} {\bibinfo  {journal} {Chemical Physics Letters}\
  }\textbf {\bibinfo {volume} {330}},\ \bibinfo {pages} {169 } (\bibinfo {year}
  {2000})}\BibitemShut {NoStop}%
\bibitem [{\citenamefont {Hino}\ \emph {et~al.}(2002)\citenamefont {Hino},
  \citenamefont {Tanimura},\ and\ \citenamefont {Ten-no}}]{hino2002transCCSD}%
  \BibitemOpen
  \bibfield  {author} {\bibinfo {author} {\bibfnamefont {O.}~\bibnamefont
  {Hino}}, \bibinfo {author} {\bibfnamefont {Y.}~\bibnamefont {Tanimura}}, \
  and\ \bibinfo {author} {\bibfnamefont {S.}~\bibnamefont {Ten-no}},\ }\href
  {\doibase https://doi.org/10.1016/S0009-2614(02)00042-8} {\bibfield
  {journal} {\bibinfo  {journal} {Chemical Physics Letters}\ }\textbf {\bibinfo
  {volume} {353}},\ \bibinfo {pages} {317 } (\bibinfo {year}
  {2002})}\BibitemShut {NoStop}%
\bibitem [{\citenamefont {Umezawa}\ and\ \citenamefont
  {Tsuneyuki}(2003)}]{umezawa2003TCmontecarlo}%
  \BibitemOpen
  \bibfield  {author} {\bibinfo {author} {\bibfnamefont {N.}~\bibnamefont
  {Umezawa}}\ and\ \bibinfo {author} {\bibfnamefont {S.}~\bibnamefont
  {Tsuneyuki}},\ }\href {\doibase 10.1063/1.1617274} {\bibfield  {journal}
  {\bibinfo  {journal} {The Journal of Chemical Physics}\ }\textbf {\bibinfo
  {volume} {119}},\ \bibinfo {pages} {10015} (\bibinfo {year} {2003})},\
  \Eprint {http://arxiv.org/abs/https://doi.org/10.1063/1.1617274}
  {https://doi.org/10.1063/1.1617274} \BibitemShut {NoStop}%
\bibitem [{\citenamefont {Umezawa}\ and\ \citenamefont
  {Tsuneyuki}(2004)}]{umezawa2004excitedTC}%
  \BibitemOpen
  \bibfield  {author} {\bibinfo {author} {\bibfnamefont {N.}~\bibnamefont
  {Umezawa}}\ and\ \bibinfo {author} {\bibfnamefont {S.}~\bibnamefont
  {Tsuneyuki}},\ }\href {\doibase 10.1063/1.1792392} {\bibfield  {journal}
  {\bibinfo  {journal} {The Journal of Chemical Physics}\ }\textbf {\bibinfo
  {volume} {121}},\ \bibinfo {pages} {7070} (\bibinfo {year} {2004})},\ \Eprint
  {http://arxiv.org/abs/https://doi.org/10.1063/1.1792392}
  {https://doi.org/10.1063/1.1792392} \BibitemShut {NoStop}%
\bibitem [{\citenamefont {Umezawa}\ \emph {et~al.}(2005)\citenamefont
  {Umezawa}, \citenamefont {Tsuneyuki}, \citenamefont {Ohno}, \citenamefont
  {Shiraishi},\ and\ \citenamefont {Chikyow}}]{umezawa2005threebody}%
  \BibitemOpen
  \bibfield  {author} {\bibinfo {author} {\bibfnamefont {N.}~\bibnamefont
  {Umezawa}}, \bibinfo {author} {\bibfnamefont {S.}~\bibnamefont {Tsuneyuki}},
  \bibinfo {author} {\bibfnamefont {T.}~\bibnamefont {Ohno}}, \bibinfo {author}
  {\bibfnamefont {K.}~\bibnamefont {Shiraishi}}, \ and\ \bibinfo {author}
  {\bibfnamefont {T.}~\bibnamefont {Chikyow}},\ }\href {\doibase
  10.1063/1.1924597} {\bibfield  {journal} {\bibinfo  {journal} {The Journal of
  Chemical Physics}\ }\textbf {\bibinfo {volume} {122}},\ \bibinfo {pages}
  {224101} (\bibinfo {year} {2005})},\ \Eprint
  {http://arxiv.org/abs/https://doi.org/10.1063/1.1924597}
  {https://doi.org/10.1063/1.1924597} \BibitemShut {NoStop}%
\bibitem [{\citenamefont {Ochi}\ \emph {et~al.}(2012)\citenamefont {Ochi},
  \citenamefont {Sodeyama}, \citenamefont {Sakuma},\ and\ \citenamefont
  {Tsuneyuki}}]{ochi2012TCperiodic}%
  \BibitemOpen
  \bibfield  {author} {\bibinfo {author} {\bibfnamefont {M.}~\bibnamefont
  {Ochi}}, \bibinfo {author} {\bibfnamefont {K.}~\bibnamefont {Sodeyama}},
  \bibinfo {author} {\bibfnamefont {R.}~\bibnamefont {Sakuma}}, \ and\ \bibinfo
  {author} {\bibfnamefont {S.}~\bibnamefont {Tsuneyuki}},\ }\href {\doibase
  10.1063/1.3689440} {\bibfield  {journal} {\bibinfo  {journal} {The Journal of
  Chemical Physics}\ }\textbf {\bibinfo {volume} {136}},\ \bibinfo {pages}
  {094108} (\bibinfo {year} {2012})},\ \Eprint
  {http://arxiv.org/abs/https://doi.org/10.1063/1.3689440}
  {https://doi.org/10.1063/1.3689440} \BibitemShut {NoStop}%
\bibitem [{\citenamefont {Luo}(2010)}]{luo2010variationalTCthrowaway}%
  \BibitemOpen
  \bibfield  {author} {\bibinfo {author} {\bibfnamefont {H.}~\bibnamefont
  {Luo}},\ }\href {\doibase 10.1063/1.3505037} {\bibfield  {journal} {\bibinfo
  {journal} {The Journal of Chemical Physics}\ }\textbf {\bibinfo {volume}
  {133}},\ \bibinfo {pages} {154109} (\bibinfo {year} {2010})},\ \Eprint
  {http://arxiv.org/abs/https://doi.org/10.1063/1.3505037}
  {https://doi.org/10.1063/1.3505037} \BibitemShut {NoStop}%
\bibitem [{\citenamefont {Luo}(2011)}]{luo2011variationalTCmcscf}%
  \BibitemOpen
  \bibfield  {author} {\bibinfo {author} {\bibfnamefont {H.}~\bibnamefont
  {Luo}},\ }\href {\doibase 10.1063/1.3607990} {\bibfield  {journal} {\bibinfo
  {journal} {The Journal of Chemical Physics}\ }\textbf {\bibinfo {volume}
  {135}},\ \bibinfo {pages} {024109} (\bibinfo {year} {2011})},\ \Eprint
  {http://arxiv.org/abs/https://doi.org/10.1063/1.3607990}
  {https://doi.org/10.1063/1.3607990} \BibitemShut {NoStop}%
\bibitem [{\citenamefont {Booth}\ \emph {et~al.}(2009)\citenamefont {Booth},
  \citenamefont {Thom},\ and\ \citenamefont {Alavi}}]{booth2009FCIQMCoriginal}%
  \BibitemOpen
  \bibfield  {author} {\bibinfo {author} {\bibfnamefont {G.~H.}\ \bibnamefont
  {Booth}}, \bibinfo {author} {\bibfnamefont {A.~J.~W.}\ \bibnamefont {Thom}},
  \ and\ \bibinfo {author} {\bibfnamefont {A.}~\bibnamefont {Alavi}},\ }\href
  {\doibase 10.1063/1.3193710} {\bibfield  {journal} {\bibinfo  {journal} {The
  Journal of Chemical Physics}\ }\textbf {\bibinfo {volume} {131}},\ \bibinfo
  {pages} {054106} (\bibinfo {year} {2009})},\ \Eprint
  {http://arxiv.org/abs/https://aip.scitation.org/doi/pdf/10.1063/1.3193710}
  {https://aip.scitation.org/doi/pdf/10.1063/1.3193710} \BibitemShut {NoStop}%
\bibitem [{\citenamefont {Jeszenszki}\ \emph {et~al.}(2018)\citenamefont
  {Jeszenszki}, \citenamefont {Luo}, \citenamefont {Alavi},\ and\ \citenamefont
  {Brand}}]{jeszenszki2018TCgases}%
  \BibitemOpen
  \bibfield  {author} {\bibinfo {author} {\bibfnamefont {P.}~\bibnamefont
  {Jeszenszki}}, \bibinfo {author} {\bibfnamefont {H.}~\bibnamefont {Luo}},
  \bibinfo {author} {\bibfnamefont {A.}~\bibnamefont {Alavi}}, \ and\ \bibinfo
  {author} {\bibfnamefont {J.}~\bibnamefont {Brand}},\ }\href {\doibase
  10.1103/PhysRevA.98.053627} {\bibfield  {journal} {\bibinfo  {journal} {Phys.
  Rev. A}\ }\textbf {\bibinfo {volume} {98}},\ \bibinfo {pages} {053627}
  (\bibinfo {year} {2018})}\BibitemShut {NoStop}%
\bibitem [{\citenamefont {Jeszenszki}\ \emph {et~al.}(2020)\citenamefont
  {Jeszenszki}, \citenamefont {Ebling}, \citenamefont {Luo}, \citenamefont
  {Alavi},\ and\ \citenamefont {Brand}}]{jeszenszki2020TCultracold}%
  \BibitemOpen
  \bibfield  {author} {\bibinfo {author} {\bibfnamefont {P.}~\bibnamefont
  {Jeszenszki}}, \bibinfo {author} {\bibfnamefont {U.}~\bibnamefont {Ebling}},
  \bibinfo {author} {\bibfnamefont {H.}~\bibnamefont {Luo}}, \bibinfo {author}
  {\bibfnamefont {A.}~\bibnamefont {Alavi}}, \ and\ \bibinfo {author}
  {\bibfnamefont {J.}~\bibnamefont {Brand}},\ }\href@noop {} {\  (\bibinfo
  {year} {2020})},\ \Eprint {http://arxiv.org/abs/arXiv:2002.05987}
  {arXiv:2002.05987} \BibitemShut {NoStop}%
\bibitem [{\citenamefont {Peruzzo}\ \emph {et~al.}(2014)\citenamefont
  {Peruzzo}, \citenamefont {McClean}, \citenamefont {Shadbolt}, \citenamefont
  {Yung}, \citenamefont {Zhou}, \citenamefont {Love}, \citenamefont
  {Aspuru-Guzik},\ and\ \citenamefont {Oâbrien}}]{peruzzo2014variational}%
  \BibitemOpen
  \bibfield  {author} {\bibinfo {author} {\bibfnamefont {A.}~\bibnamefont
  {Peruzzo}}, \bibinfo {author} {\bibfnamefont {J.}~\bibnamefont {McClean}},
  \bibinfo {author} {\bibfnamefont {P.}~\bibnamefont {Shadbolt}}, \bibinfo
  {author} {\bibfnamefont {M.-H.}\ \bibnamefont {Yung}}, \bibinfo {author}
  {\bibfnamefont {X.-Q.}\ \bibnamefont {Zhou}}, \bibinfo {author}
  {\bibfnamefont {P.~J.}\ \bibnamefont {Love}}, \bibinfo {author}
  {\bibfnamefont {A.}~\bibnamefont {Aspuru-Guzik}}, \ and\ \bibinfo {author}
  {\bibfnamefont {J.~L.}\ \bibnamefont {Oâbrien}},\ }\href@noop {}
  {\bibfield  {journal} {\bibinfo  {journal} {Nature communications}\ }\textbf
  {\bibinfo {volume} {5}} (\bibinfo {year} {2014})}\BibitemShut {NoStop}%
\bibitem [{\citenamefont {McClean}\ \emph {et~al.}(2016)\citenamefont
  {McClean}, \citenamefont {Romero}, \citenamefont {Babbush},\ and\
  \citenamefont {Aspuru-Guzik}}]{VQETheoryNJP}%
  \BibitemOpen
  \bibfield  {author} {\bibinfo {author} {\bibfnamefont {J.~R.}\ \bibnamefont
  {McClean}}, \bibinfo {author} {\bibfnamefont {J.}~\bibnamefont {Romero}},
  \bibinfo {author} {\bibfnamefont {R.}~\bibnamefont {Babbush}}, \ and\
  \bibinfo {author} {\bibfnamefont {A.}~\bibnamefont {Aspuru-Guzik}},\ }\href
  {http://stacks.iop.org/1367-2630/18/i=2/a=023023} {\bibfield  {journal}
  {\bibinfo  {journal} {New Journal of Physics}\ }\textbf {\bibinfo {volume}
  {18}},\ \bibinfo {pages} {023023} (\bibinfo {year} {2016})}\BibitemShut
  {NoStop}%
\bibitem [{\citenamefont {Kitaev}(1995)}]{kitaev1995phase}%
  \BibitemOpen
  \bibfield  {author} {\bibinfo {author} {\bibfnamefont {A.~Y.}\ \bibnamefont
  {Kitaev}},\ }\href@noop {} {\bibfield  {journal} {\bibinfo  {journal}
  {Preprint at http://arxiv. org/abs/quant-ph/9511026}\ } (\bibinfo {year}
  {1995})}\BibitemShut {NoStop}%
\bibitem [{\citenamefont {Bauman}\ \emph
  {et~al.}(2019{\natexlab{a}})\citenamefont {Bauman}, \citenamefont {Bylaska},
  \citenamefont {Krishnamoorthy}, \citenamefont {Low}, \citenamefont {Wiebe},
  \citenamefont {Granade}, \citenamefont {Roetteler}, \citenamefont {Troyer},\
  and\ \citenamefont {Kowalski}}]{bauman2019downfolding}%
  \BibitemOpen
  \bibfield  {author} {\bibinfo {author} {\bibfnamefont {N.~P.}\ \bibnamefont
  {Bauman}}, \bibinfo {author} {\bibfnamefont {E.~J.}\ \bibnamefont {Bylaska}},
  \bibinfo {author} {\bibfnamefont {S.}~\bibnamefont {Krishnamoorthy}},
  \bibinfo {author} {\bibfnamefont {G.~H.}\ \bibnamefont {Low}}, \bibinfo
  {author} {\bibfnamefont {N.}~\bibnamefont {Wiebe}}, \bibinfo {author}
  {\bibfnamefont {C.~E.}\ \bibnamefont {Granade}}, \bibinfo {author}
  {\bibfnamefont {M.}~\bibnamefont {Roetteler}}, \bibinfo {author}
  {\bibfnamefont {M.}~\bibnamefont {Troyer}}, \ and\ \bibinfo {author}
  {\bibfnamefont {K.}~\bibnamefont {Kowalski}},\ }\href {\doibase
  10.1063/1.5094643} {\bibfield  {journal} {\bibinfo  {journal} {The Journal of
  Chemical Physics}\ }\textbf {\bibinfo {volume} {151}},\ \bibinfo {pages}
  {014107} (\bibinfo {year} {2019}{\natexlab{a}})},\ \Eprint
  {http://arxiv.org/abs/https://doi.org/10.1063/1.5094643}
  {https://doi.org/10.1063/1.5094643} \BibitemShut {NoStop}%
\bibitem [{\citenamefont {Bauman}\ \emph
  {et~al.}(2019{\natexlab{b}})\citenamefont {Bauman}, \citenamefont {Low},\
  and\ \citenamefont {Kowalski}}]{bauman2019downfoldingexcited}%
  \BibitemOpen
  \bibfield  {author} {\bibinfo {author} {\bibfnamefont {N.~P.}\ \bibnamefont
  {Bauman}}, \bibinfo {author} {\bibfnamefont {G.~H.}\ \bibnamefont {Low}}, \
  and\ \bibinfo {author} {\bibfnamefont {K.}~\bibnamefont {Kowalski}},\ }\href
  {\doibase 10.1063/1.5128103} {\bibfield  {journal} {\bibinfo  {journal} {The
  Journal of Chemical Physics}\ }\textbf {\bibinfo {volume} {151}},\ \bibinfo
  {pages} {234114} (\bibinfo {year} {2019}{\natexlab{b}})},\ \Eprint
  {http://arxiv.org/abs/https://doi.org/10.1063/1.5128103}
  {https://doi.org/10.1063/1.5128103} \BibitemShut {NoStop}%
\bibitem [{\citenamefont {Takeshita}\ \emph {et~al.}(2020)\citenamefont
  {Takeshita}, \citenamefont {Rubin}, \citenamefont {Jiang}, \citenamefont
  {Lee}, \citenamefont {Babbush},\ and\ \citenamefont
  {McClean}}]{takeshita2019virtualorbs}%
  \BibitemOpen
  \bibfield  {author} {\bibinfo {author} {\bibfnamefont {T.}~\bibnamefont
  {Takeshita}}, \bibinfo {author} {\bibfnamefont {N.~C.}\ \bibnamefont
  {Rubin}}, \bibinfo {author} {\bibfnamefont {Z.}~\bibnamefont {Jiang}},
  \bibinfo {author} {\bibfnamefont {E.}~\bibnamefont {Lee}}, \bibinfo {author}
  {\bibfnamefont {R.}~\bibnamefont {Babbush}}, \ and\ \bibinfo {author}
  {\bibfnamefont {J.~R.}\ \bibnamefont {McClean}},\ }\href {\doibase
  10.1103/PhysRevX.10.011004} {\bibfield  {journal} {\bibinfo  {journal} {Phys.
  Rev. X}\ }\textbf {\bibinfo {volume} {10}},\ \bibinfo {pages} {011004}
  (\bibinfo {year} {2020})}\BibitemShut {NoStop}%
\bibitem [{\citenamefont {Motta}\ \emph {et~al.}(2019)\citenamefont {Motta},
  \citenamefont {Sun}, \citenamefont {Tan}, \citenamefont {O'Rourke},
  \citenamefont {Ye}, \citenamefont {Minnich}, \citenamefont {Brand{\~{a}}o},\
  and\ \citenamefont {Chan}}]{motta2019imaginary}%
  \BibitemOpen
  \bibfield  {author} {\bibinfo {author} {\bibfnamefont {M.}~\bibnamefont
  {Motta}}, \bibinfo {author} {\bibfnamefont {C.}~\bibnamefont {Sun}}, \bibinfo
  {author} {\bibfnamefont {A.~T.~K.}\ \bibnamefont {Tan}}, \bibinfo {author}
  {\bibfnamefont {M.~J.}\ \bibnamefont {O'Rourke}}, \bibinfo {author}
  {\bibfnamefont {E.}~\bibnamefont {Ye}}, \bibinfo {author} {\bibfnamefont
  {A.~J.}\ \bibnamefont {Minnich}}, \bibinfo {author} {\bibfnamefont {F.~G.
  S.~L.}\ \bibnamefont {Brand{\~{a}}o}}, \ and\ \bibinfo {author}
  {\bibfnamefont {G.~K.-L.}\ \bibnamefont {Chan}},\ }\href {\doibase
  10.1038/s41567-019-0704-4} {\bibfield  {journal} {\bibinfo  {journal} {Nature
  Physics}\ }\textbf {\bibinfo {volume} {16}},\ \bibinfo {pages} {205}
  (\bibinfo {year} {2019})}\BibitemShut {NoStop}%
\bibitem [{\citenamefont {Hackl}\ \emph {et~al.}(2020)\citenamefont {Hackl},
  \citenamefont {Guaita}, \citenamefont {Shi}, \citenamefont {Haegeman},
  \citenamefont {Demler},\ and\ \citenamefont
  {Cirac}}]{hackl2020GeometryImagTime}%
  \BibitemOpen
  \bibfield  {author} {\bibinfo {author} {\bibfnamefont {L.}~\bibnamefont
  {Hackl}}, \bibinfo {author} {\bibfnamefont {T.}~\bibnamefont {Guaita}},
  \bibinfo {author} {\bibfnamefont {T.}~\bibnamefont {Shi}}, \bibinfo {author}
  {\bibfnamefont {J.}~\bibnamefont {Haegeman}}, \bibinfo {author}
  {\bibfnamefont {E.}~\bibnamefont {Demler}}, \ and\ \bibinfo {author}
  {\bibfnamefont {I.}~\bibnamefont {Cirac}},\ }\href@noop {} {\  (\bibinfo
  {year} {2020})},\ \Eprint {http://arxiv.org/abs/arXiv:2004.01015}
  {arXiv:2004.01015} \BibitemShut {NoStop}%
\bibitem [{\citenamefont {Li}\ and\ \citenamefont {Benjamin}(2017)}]{Li2017}%
  \BibitemOpen
  \bibfield  {author} {\bibinfo {author} {\bibfnamefont {Y.}~\bibnamefont
  {Li}}\ and\ \bibinfo {author} {\bibfnamefont {S.~C.}\ \bibnamefont
  {Benjamin}},\ }\href {\doibase 10.1103/PhysRevX.7.021050} {\bibfield
  {journal} {\bibinfo  {journal} {Phys. Rev. X}\ }\textbf {\bibinfo {volume}
  {7}},\ \bibinfo {pages} {021050} (\bibinfo {year} {2017})}\BibitemShut
  {NoStop}%
\bibitem [{\citenamefont {Yuan}\ \emph {et~al.}(2019)\citenamefont {Yuan},
  \citenamefont {Endo}, \citenamefont {Zhao}, \citenamefont {Li},\ and\
  \citenamefont {Benjamin}}]{yuan2018variationaltheory}%
  \BibitemOpen
  \bibfield  {author} {\bibinfo {author} {\bibfnamefont {X.}~\bibnamefont
  {Yuan}}, \bibinfo {author} {\bibfnamefont {S.}~\bibnamefont {Endo}}, \bibinfo
  {author} {\bibfnamefont {Q.}~\bibnamefont {Zhao}}, \bibinfo {author}
  {\bibfnamefont {Y.}~\bibnamefont {Li}}, \ and\ \bibinfo {author}
  {\bibfnamefont {S.~C.}\ \bibnamefont {Benjamin}},\ }\href {\doibase
  10.22331/q-2019-10-07-191} {\bibfield  {journal} {\bibinfo  {journal}
  {{Quantum}}\ }\textbf {\bibinfo {volume} {3}},\ \bibinfo {pages} {191}
  (\bibinfo {year} {2019})}\BibitemShut {NoStop}%
\bibitem [{\citenamefont {Endo}\ \emph {et~al.}(2018)\citenamefont {Endo},
  \citenamefont {Li}, \citenamefont {Benjamin},\ and\ \citenamefont
  {Yuan}}]{endo2018variational}%
  \BibitemOpen
  \bibfield  {author} {\bibinfo {author} {\bibfnamefont {S.}~\bibnamefont
  {Endo}}, \bibinfo {author} {\bibfnamefont {Y.}~\bibnamefont {Li}}, \bibinfo
  {author} {\bibfnamefont {S.}~\bibnamefont {Benjamin}}, \ and\ \bibinfo
  {author} {\bibfnamefont {X.}~\bibnamefont {Yuan}},\ }\href@noop {} {\bibfield
   {journal} {\bibinfo  {journal} {arXiv preprint arXiv:1812.08778}\ }
  (\bibinfo {year} {2018})}\BibitemShut {NoStop}%
\bibitem [{\citenamefont {Koczor}\ and\ \citenamefont
  {Benjamin}(2019)}]{koczor2019NaturalGrad}%
  \BibitemOpen
  \bibfield  {author} {\bibinfo {author} {\bibfnamefont {B.}~\bibnamefont
  {Koczor}}\ and\ \bibinfo {author} {\bibfnamefont {S.~C.}\ \bibnamefont
  {Benjamin}},\ }\href@noop {} {\  (\bibinfo {year} {2019})},\ \Eprint
  {http://arxiv.org/abs/arXiv:1912.08660} {arXiv:1912.08660} \BibitemShut
  {NoStop}%
\bibitem [{\citenamefont {Jones}\ \emph {et~al.}(2019)\citenamefont {Jones},
  \citenamefont {Endo}, \citenamefont {McArdle}, \citenamefont {Yuan},\ and\
  \citenamefont {Benjamin}}]{jones2019ImagExcited}%
  \BibitemOpen
  \bibfield  {author} {\bibinfo {author} {\bibfnamefont {T.}~\bibnamefont
  {Jones}}, \bibinfo {author} {\bibfnamefont {S.}~\bibnamefont {Endo}},
  \bibinfo {author} {\bibfnamefont {S.}~\bibnamefont {McArdle}}, \bibinfo
  {author} {\bibfnamefont {X.}~\bibnamefont {Yuan}}, \ and\ \bibinfo {author}
  {\bibfnamefont {S.~C.}\ \bibnamefont {Benjamin}},\ }\href {\doibase
  10.1103/PhysRevA.99.062304} {\bibfield  {journal} {\bibinfo  {journal} {Phys.
  Rev. A}\ }\textbf {\bibinfo {volume} {99}},\ \bibinfo {pages} {062304}
  (\bibinfo {year} {2019})}\BibitemShut {NoStop}%
\bibitem [{\citenamefont {Jones}\ and\ \citenamefont
  {Benjamin}(2018)}]{jones2018compiling}%
  \BibitemOpen
  \bibfield  {author} {\bibinfo {author} {\bibfnamefont {T.}~\bibnamefont
  {Jones}}\ and\ \bibinfo {author} {\bibfnamefont {S.~C.}\ \bibnamefont
  {Benjamin}},\ }\href@noop {} {\  (\bibinfo {year} {2018})},\ \Eprint
  {http://arxiv.org/abs/arXiv:1811.03147} {arXiv:1811.03147} \BibitemShut
  {NoStop}%
\bibitem [{\citenamefont {Xu}\ \emph {et~al.}(2019{\natexlab{a}})\citenamefont
  {Xu}, \citenamefont {Benjamin},\ and\ \citenamefont
  {Yuan}}]{xu2019compiling}%
  \BibitemOpen
  \bibfield  {author} {\bibinfo {author} {\bibfnamefont {X.}~\bibnamefont
  {Xu}}, \bibinfo {author} {\bibfnamefont {S.~C.}\ \bibnamefont {Benjamin}}, \
  and\ \bibinfo {author} {\bibfnamefont {X.}~\bibnamefont {Yuan}},\ }\href@noop
  {} {\  (\bibinfo {year} {2019}{\natexlab{a}})},\ \Eprint
  {http://arxiv.org/abs/arXiv:1911.05759} {arXiv:1911.05759} \BibitemShut
  {NoStop}%
\bibitem [{\citenamefont {Zoufal}\ \emph {et~al.}(2020)\citenamefont {Zoufal},
  \citenamefont {Lucchi},\ and\ \citenamefont
  {Woerner}}]{zoufal2020ImagTimeBoltzmann}%
  \BibitemOpen
  \bibfield  {author} {\bibinfo {author} {\bibfnamefont {C.}~\bibnamefont
  {Zoufal}}, \bibinfo {author} {\bibfnamefont {A.}~\bibnamefont {Lucchi}}, \
  and\ \bibinfo {author} {\bibfnamefont {S.}~\bibnamefont {Woerner}},\
  }\href@noop {} {\  (\bibinfo {year} {2020})},\ \Eprint
  {http://arxiv.org/abs/arXiv:2006.06004} {arXiv:2006.06004} \BibitemShut
  {NoStop}%
\bibitem [{\citenamefont {Liu}\ and\ \citenamefont
  {Xin}(2020)}]{liu2020ImagTimeFieldTheories}%
  \BibitemOpen
  \bibfield  {author} {\bibinfo {author} {\bibfnamefont {J.}~\bibnamefont
  {Liu}}\ and\ \bibinfo {author} {\bibfnamefont {Y.}~\bibnamefont {Xin}},\
  }\href@noop {} {\  (\bibinfo {year} {2020})},\ \Eprint
  {http://arxiv.org/abs/arXiv:2004.13234} {arXiv:2004.13234} \BibitemShut
  {NoStop}%
\bibitem [{\citenamefont {Xu}\ \emph {et~al.}(2019{\natexlab{b}})\citenamefont
  {Xu}, \citenamefont {Sun}, \citenamefont {Endo}, \citenamefont {Li},
  \citenamefont {Benjamin},\ and\ \citenamefont {Yuan}}]{xu2019LinearAlgebra}%
  \BibitemOpen
  \bibfield  {author} {\bibinfo {author} {\bibfnamefont {X.}~\bibnamefont
  {Xu}}, \bibinfo {author} {\bibfnamefont {J.}~\bibnamefont {Sun}}, \bibinfo
  {author} {\bibfnamefont {S.}~\bibnamefont {Endo}}, \bibinfo {author}
  {\bibfnamefont {Y.}~\bibnamefont {Li}}, \bibinfo {author} {\bibfnamefont
  {S.~C.}\ \bibnamefont {Benjamin}}, \ and\ \bibinfo {author} {\bibfnamefont
  {X.}~\bibnamefont {Yuan}},\ }\href@noop {} {\  (\bibinfo {year}
  {2019}{\natexlab{b}})},\ \Eprint {http://arxiv.org/abs/arXiv:1909.03898}
  {arXiv:1909.03898} \BibitemShut {NoStop}%
\bibitem [{\citenamefont {Huang}\ \emph {et~al.}(2019)\citenamefont {Huang},
  \citenamefont {Bharti},\ and\ \citenamefont
  {Rebentrost}}]{huang2019LinearAlgebra}%
  \BibitemOpen
  \bibfield  {author} {\bibinfo {author} {\bibfnamefont {H.-Y.}\ \bibnamefont
  {Huang}}, \bibinfo {author} {\bibfnamefont {K.}~\bibnamefont {Bharti}}, \
  and\ \bibinfo {author} {\bibfnamefont {P.}~\bibnamefont {Rebentrost}},\
  }\href@noop {} {\  (\bibinfo {year} {2019})},\ \Eprint
  {http://arxiv.org/abs/arXiv:1909.07344} {arXiv:1909.07344} \BibitemShut
  {NoStop}%
\bibitem [{\citenamefont {Fontanela}\ \emph {et~al.}(2019)\citenamefont
  {Fontanela}, \citenamefont {Jacquier},\ and\ \citenamefont
  {Oumgari}}]{fontanela2019ImagTimeFinance}%
  \BibitemOpen
  \bibfield  {author} {\bibinfo {author} {\bibfnamefont {F.}~\bibnamefont
  {Fontanela}}, \bibinfo {author} {\bibfnamefont {A.}~\bibnamefont {Jacquier}},
  \ and\ \bibinfo {author} {\bibfnamefont {M.}~\bibnamefont {Oumgari}},\
  }\href@noop {} {\  (\bibinfo {year} {2019})},\ \Eprint
  {http://arxiv.org/abs/arXiv:1912.02753} {arXiv:1912.02753} \BibitemShut
  {NoStop}%
\bibitem [{\citenamefont {McLachlan}(1964)}]{mclachlan1964variational}%
  \BibitemOpen
  \bibfield  {author} {\bibinfo {author} {\bibfnamefont {A.}~\bibnamefont
  {McLachlan}},\ }\href {\doibase 10.1080/00268976400100041} {\bibfield
  {journal} {\bibinfo  {journal} {Molecular Physics}\ }\textbf {\bibinfo
  {volume} {8}},\ \bibinfo {pages} {39} (\bibinfo {year} {1964})},\ \Eprint
  {http://arxiv.org/abs/https://doi.org/10.1080/00268976400100041}
  {https://doi.org/10.1080/00268976400100041} \BibitemShut {NoStop}%
\bibitem [{\citenamefont {van Straaten}\ and\ \citenamefont
  {Koczor}(2020)}]{VanStraaten2020Measurement}%
  \BibitemOpen
  \bibfield  {author} {\bibinfo {author} {\bibfnamefont {B.}~\bibnamefont {van
  Straaten}}\ and\ \bibinfo {author} {\bibfnamefont {B.}~\bibnamefont
  {Koczor}},\ }\href@noop {} {\  (\bibinfo {year} {2020})},\ \Eprint
  {http://arxiv.org/abs/arXiv:2005.05172} {arXiv:2005.05172} \BibitemShut
  {NoStop}%
\bibitem [{\citenamefont {Mitarai}\ and\ \citenamefont
  {Fujii}(2019)}]{mitarai2018imaginaryindirect}%
  \BibitemOpen
  \bibfield  {author} {\bibinfo {author} {\bibfnamefont {K.}~\bibnamefont
  {Mitarai}}\ and\ \bibinfo {author} {\bibfnamefont {K.}~\bibnamefont
  {Fujii}},\ }\href {\doibase 10.1103/PhysRevResearch.1.013006} {\bibfield
  {journal} {\bibinfo  {journal} {Phys. Rev. Research}\ }\textbf {\bibinfo
  {volume} {1}},\ \bibinfo {pages} {013006} (\bibinfo {year}
  {2019})}\BibitemShut {NoStop}%
\bibitem [{\citenamefont {Schuld}\ \emph {et~al.}(2019)\citenamefont {Schuld},
  \citenamefont {Bergholm}, \citenamefont {Gogolin}, \citenamefont {Izaac},\
  and\ \citenamefont {Killoran}}]{schuld2019gradients}%
  \BibitemOpen
  \bibfield  {author} {\bibinfo {author} {\bibfnamefont {M.}~\bibnamefont
  {Schuld}}, \bibinfo {author} {\bibfnamefont {V.}~\bibnamefont {Bergholm}},
  \bibinfo {author} {\bibfnamefont {C.}~\bibnamefont {Gogolin}}, \bibinfo
  {author} {\bibfnamefont {J.}~\bibnamefont {Izaac}}, \ and\ \bibinfo {author}
  {\bibfnamefont {N.}~\bibnamefont {Killoran}},\ }\href {\doibase
  10.1103/PhysRevA.99.032331} {\bibfield  {journal} {\bibinfo  {journal} {Phys.
  Rev. A}\ }\textbf {\bibinfo {volume} {99}},\ \bibinfo {pages} {032331}
  (\bibinfo {year} {2019})}\BibitemShut {NoStop}%
\bibitem [{\citenamefont {Google}(2020)}]{google2020cirq}%
  \BibitemOpen
  \bibfield  {author} {\bibinfo {author} {\bibnamefont {Google}},\ }\href@noop
  {} {\enquote {\bibinfo {title} {Cirq},}\ }\bibinfo {howpublished}
  {\url{https://github.com/quantumlib/Cirq}} (\bibinfo {year}
  {2020})\BibitemShut {NoStop}%
\bibitem [{\citenamefont {Dagotto}(1994)}]{dagotto1994correlatedelectrons}%
  \BibitemOpen
  \bibfield  {author} {\bibinfo {author} {\bibfnamefont {E.}~\bibnamefont
  {Dagotto}},\ }\href {\doibase 10.1103/RevModPhys.66.763} {\bibfield
  {journal} {\bibinfo  {journal} {Rev. Mod. Phys.}\ }\textbf {\bibinfo {volume}
  {66}},\ \bibinfo {pages} {763} (\bibinfo {year} {1994})}\BibitemShut
  {NoStop}%
\bibitem [{\citenamefont {Anderson}(2002)}]{anderson2002hubbard}%
  \BibitemOpen
  \bibfield  {author} {\bibinfo {author} {\bibfnamefont {P.~W.}\ \bibnamefont
  {Anderson}},\ }\href {\doibase 10.1238/physica.topical.102a00010} {\bibfield
  {journal} {\bibinfo  {journal} {Physica Scripta}\ }\textbf {\bibinfo {volume}
  {T102}},\ \bibinfo {pages} {10} (\bibinfo {year} {2002})}\BibitemShut
  {NoStop}%
\bibitem [{\citenamefont {LeBlanc}\ \emph {et~al.}(2015)\citenamefont
  {LeBlanc}, \citenamefont {Antipov}, \citenamefont {Becca}, \citenamefont
  {Bulik}, \citenamefont {Chan}, \citenamefont {Chung}, \citenamefont {Deng},
  \citenamefont {Ferrero}, \citenamefont {Henderson}, \citenamefont
  {Jim\'enez-Hoyos}, \citenamefont {Kozik}, \citenamefont {Liu}, \citenamefont
  {Millis}, \citenamefont {Prokof'ev}, \citenamefont {Qin}, \citenamefont
  {Scuseria}, \citenamefont {Shi}, \citenamefont {Svistunov}, \citenamefont
  {Tocchio}, \citenamefont {Tupitsyn}, \citenamefont {White}, \citenamefont
  {Zhang}, \citenamefont {Zheng}, \citenamefont {Zhu},\ and\ \citenamefont
  {Gull}}]{simonscollab2015hubbard}%
  \BibitemOpen
  \bibfield  {author} {\bibinfo {author} {\bibfnamefont {J.~P.~F.}\
  \bibnamefont {LeBlanc}}, \bibinfo {author} {\bibfnamefont {A.~E.}\
  \bibnamefont {Antipov}}, \bibinfo {author} {\bibfnamefont {F.}~\bibnamefont
  {Becca}}, \bibinfo {author} {\bibfnamefont {I.~W.}\ \bibnamefont {Bulik}},
  \bibinfo {author} {\bibfnamefont {G.~K.-L.}\ \bibnamefont {Chan}}, \bibinfo
  {author} {\bibfnamefont {C.-M.}\ \bibnamefont {Chung}}, \bibinfo {author}
  {\bibfnamefont {Y.}~\bibnamefont {Deng}}, \bibinfo {author} {\bibfnamefont
  {M.}~\bibnamefont {Ferrero}}, \bibinfo {author} {\bibfnamefont {T.~M.}\
  \bibnamefont {Henderson}}, \bibinfo {author} {\bibfnamefont {C.~A.}\
  \bibnamefont {Jim\'enez-Hoyos}}, \bibinfo {author} {\bibfnamefont
  {E.}~\bibnamefont {Kozik}}, \bibinfo {author} {\bibfnamefont {X.-W.}\
  \bibnamefont {Liu}}, \bibinfo {author} {\bibfnamefont {A.~J.}\ \bibnamefont
  {Millis}}, \bibinfo {author} {\bibfnamefont {N.~V.}\ \bibnamefont
  {Prokof'ev}}, \bibinfo {author} {\bibfnamefont {M.}~\bibnamefont {Qin}},
  \bibinfo {author} {\bibfnamefont {G.~E.}\ \bibnamefont {Scuseria}}, \bibinfo
  {author} {\bibfnamefont {H.}~\bibnamefont {Shi}}, \bibinfo {author}
  {\bibfnamefont {B.~V.}\ \bibnamefont {Svistunov}}, \bibinfo {author}
  {\bibfnamefont {L.~F.}\ \bibnamefont {Tocchio}}, \bibinfo {author}
  {\bibfnamefont {I.~S.}\ \bibnamefont {Tupitsyn}}, \bibinfo {author}
  {\bibfnamefont {S.~R.}\ \bibnamefont {White}}, \bibinfo {author}
  {\bibfnamefont {S.}~\bibnamefont {Zhang}}, \bibinfo {author} {\bibfnamefont
  {B.-X.}\ \bibnamefont {Zheng}}, \bibinfo {author} {\bibfnamefont
  {Z.}~\bibnamefont {Zhu}}, \ and\ \bibinfo {author} {\bibfnamefont
  {E.}~\bibnamefont {Gull}} (\bibinfo {collaboration} {Simons Collaboration on
  the Many-Electron Problem}),\ }\href {\doibase 10.1103/PhysRevX.5.041041}
  {\bibfield  {journal} {\bibinfo  {journal} {Phys. Rev. X}\ }\textbf {\bibinfo
  {volume} {5}},\ \bibinfo {pages} {041041} (\bibinfo {year}
  {2015})}\BibitemShut {NoStop}%
\bibitem [{\citenamefont {McClean}\ \emph
  {et~al.}(2017{\natexlab{a}})\citenamefont {McClean}, \citenamefont
  {Kivlichan}, \citenamefont {Sung}, \citenamefont {Steiger}, \citenamefont
  {Cao}, \citenamefont {Dai}, \citenamefont {Fried}, \citenamefont {Gidney},
  \citenamefont {Gimby}, \citenamefont {Gokhale}, \citenamefont {HÃ¤ner},
  \citenamefont {Hardikar}, \citenamefont {HavlÃ­Äek}, \citenamefont
  {Huang}, \citenamefont {Izaac}, \citenamefont {Jiang}, \citenamefont {Liu},
  \citenamefont {Neeley}, \citenamefont {O'Brien}, \citenamefont {Ozfidan},
  \citenamefont {Radin}, \citenamefont {Romero}, \citenamefont {Rubin},
  \citenamefont {Sawaya}, \citenamefont {Setia}, \citenamefont {Sim},
  \citenamefont {Steudtner}, \citenamefont {Sun}, \citenamefont {Sun},
  \citenamefont {Zhang},\ and\ \citenamefont
  {Babbush}}]{mcclean2017openfermion}%
  \BibitemOpen
  \bibfield  {author} {\bibinfo {author} {\bibfnamefont {J.~R.}\ \bibnamefont
  {McClean}}, \bibinfo {author} {\bibfnamefont {I.~D.}\ \bibnamefont
  {Kivlichan}}, \bibinfo {author} {\bibfnamefont {K.~J.}\ \bibnamefont {Sung}},
  \bibinfo {author} {\bibfnamefont {D.~S.}\ \bibnamefont {Steiger}}, \bibinfo
  {author} {\bibfnamefont {Y.}~\bibnamefont {Cao}}, \bibinfo {author}
  {\bibfnamefont {C.}~\bibnamefont {Dai}}, \bibinfo {author} {\bibfnamefont
  {E.~S.}\ \bibnamefont {Fried}}, \bibinfo {author} {\bibfnamefont
  {C.}~\bibnamefont {Gidney}}, \bibinfo {author} {\bibfnamefont
  {B.}~\bibnamefont {Gimby}}, \bibinfo {author} {\bibfnamefont
  {P.}~\bibnamefont {Gokhale}}, \bibinfo {author} {\bibfnamefont
  {T.}~\bibnamefont {HÃ¤ner}}, \bibinfo {author} {\bibfnamefont
  {T.}~\bibnamefont {Hardikar}}, \bibinfo {author} {\bibfnamefont
  {V.}~\bibnamefont {HavlÃ­Äek}}, \bibinfo {author} {\bibfnamefont
  {C.}~\bibnamefont {Huang}}, \bibinfo {author} {\bibfnamefont
  {J.}~\bibnamefont {Izaac}}, \bibinfo {author} {\bibfnamefont
  {Z.}~\bibnamefont {Jiang}}, \bibinfo {author} {\bibfnamefont
  {X.}~\bibnamefont {Liu}}, \bibinfo {author} {\bibfnamefont {M.}~\bibnamefont
  {Neeley}}, \bibinfo {author} {\bibfnamefont {T.}~\bibnamefont {O'Brien}},
  \bibinfo {author} {\bibfnamefont {I.}~\bibnamefont {Ozfidan}}, \bibinfo
  {author} {\bibfnamefont {M.~D.}\ \bibnamefont {Radin}}, \bibinfo {author}
  {\bibfnamefont {J.}~\bibnamefont {Romero}}, \bibinfo {author} {\bibfnamefont
  {N.}~\bibnamefont {Rubin}}, \bibinfo {author} {\bibfnamefont {N.~P.~D.}\
  \bibnamefont {Sawaya}}, \bibinfo {author} {\bibfnamefont {K.}~\bibnamefont
  {Setia}}, \bibinfo {author} {\bibfnamefont {S.}~\bibnamefont {Sim}}, \bibinfo
  {author} {\bibfnamefont {M.}~\bibnamefont {Steudtner}}, \bibinfo {author}
  {\bibfnamefont {Q.}~\bibnamefont {Sun}}, \bibinfo {author} {\bibfnamefont
  {W.}~\bibnamefont {Sun}}, \bibinfo {author} {\bibfnamefont {F.}~\bibnamefont
  {Zhang}}, \ and\ \bibinfo {author} {\bibfnamefont {R.}~\bibnamefont
  {Babbush}},\ }\href@noop {} {\bibfield  {journal} {\bibinfo  {journal}
  {arXiv:1710.07629}\ } (\bibinfo {year} {2017}{\natexlab{a}})}\BibitemShut
  {NoStop}%
\bibitem [{\citenamefont {Wecker}\ \emph
  {et~al.}(2015{\natexlab{a}})\citenamefont {Wecker}, \citenamefont
  {Hastings},\ and\ \citenamefont {Troyer}}]{PhysRevA.92.042303}%
  \BibitemOpen
  \bibfield  {author} {\bibinfo {author} {\bibfnamefont {D.}~\bibnamefont
  {Wecker}}, \bibinfo {author} {\bibfnamefont {M.~B.}\ \bibnamefont
  {Hastings}}, \ and\ \bibinfo {author} {\bibfnamefont {M.}~\bibnamefont
  {Troyer}},\ }\href {\doibase 10.1103/PhysRevA.92.042303} {\bibfield
  {journal} {\bibinfo  {journal} {Phys. Rev. A}\ }\textbf {\bibinfo {volume}
  {92}},\ \bibinfo {pages} {042303} (\bibinfo {year}
  {2015}{\natexlab{a}})}\BibitemShut {NoStop}%
\bibitem [{\citenamefont {Cai}(2019)}]{cai2019hubbard}%
  \BibitemOpen
  \bibfield  {author} {\bibinfo {author} {\bibfnamefont {Z.}~\bibnamefont
  {Cai}},\ }\href@noop {} {\  (\bibinfo {year} {2019})},\ \Eprint
  {http://arxiv.org/abs/arXiv:1910.02719} {arXiv:1910.02719} \BibitemShut
  {NoStop}%
\bibitem [{\citenamefont {Reiner}\ \emph {et~al.}(2019)\citenamefont {Reiner},
  \citenamefont {Wilhelm-Mauch}, \citenamefont {Sch\"{o}n},\ and\ \citenamefont
  {Marthaler}}]{reiner2018hubbardground}%
  \BibitemOpen
  \bibfield  {author} {\bibinfo {author} {\bibfnamefont {J.-M.}\ \bibnamefont
  {Reiner}}, \bibinfo {author} {\bibfnamefont {F.}~\bibnamefont
  {Wilhelm-Mauch}}, \bibinfo {author} {\bibfnamefont {G.}~\bibnamefont
  {Sch\"{o}n}}, \ and\ \bibinfo {author} {\bibfnamefont {M.}~\bibnamefont
  {Marthaler}},\ }\href {\doibase 10.1088/2058-9565/ab1e85} {\bibfield
  {journal} {\bibinfo  {journal} {Quantum Science and Technology}\ }\textbf
  {\bibinfo {volume} {4}},\ \bibinfo {pages} {035005} (\bibinfo {year}
  {2019})}\BibitemShut {NoStop}%
\bibitem [{\citenamefont {Cade}\ \emph {et~al.}(2019)\citenamefont {Cade},
  \citenamefont {Mineh}, \citenamefont {Montanaro},\ and\ \citenamefont
  {Stanisic}}]{cade2019fermihubbard}%
  \BibitemOpen
  \bibfield  {author} {\bibinfo {author} {\bibfnamefont {C.}~\bibnamefont
  {Cade}}, \bibinfo {author} {\bibfnamefont {L.}~\bibnamefont {Mineh}},
  \bibinfo {author} {\bibfnamefont {A.}~\bibnamefont {Montanaro}}, \ and\
  \bibinfo {author} {\bibfnamefont {S.}~\bibnamefont {Stanisic}},\ }\href@noop
  {} {\  (\bibinfo {year} {2019})},\ \Eprint
  {http://arxiv.org/abs/arXiv:1912.06007} {arXiv:1912.06007} \BibitemShut
  {NoStop}%
\bibitem [{\citenamefont {Wecker}\ \emph
  {et~al.}(2015{\natexlab{b}})\citenamefont {Wecker}, \citenamefont {Hastings},
  \citenamefont {Wiebe}, \citenamefont {Clark}, \citenamefont {Nayak},\ and\
  \citenamefont {Troyer}}]{wecker2015hubbard}%
  \BibitemOpen
  \bibfield  {author} {\bibinfo {author} {\bibfnamefont {D.}~\bibnamefont
  {Wecker}}, \bibinfo {author} {\bibfnamefont {M.~B.}\ \bibnamefont
  {Hastings}}, \bibinfo {author} {\bibfnamefont {N.}~\bibnamefont {Wiebe}},
  \bibinfo {author} {\bibfnamefont {B.~K.}\ \bibnamefont {Clark}}, \bibinfo
  {author} {\bibfnamefont {C.}~\bibnamefont {Nayak}}, \ and\ \bibinfo {author}
  {\bibfnamefont {M.}~\bibnamefont {Troyer}},\ }\href {\doibase
  10.1103/PhysRevA.92.062318} {\bibfield  {journal} {\bibinfo  {journal} {Phys.
  Rev. A}\ }\textbf {\bibinfo {volume} {92}},\ \bibinfo {pages} {062318}
  (\bibinfo {year} {2015}{\natexlab{b}})}\BibitemShut {NoStop}%
\bibitem [{\citenamefont {Kivlichan}\ \emph {et~al.}(2018)\citenamefont
  {Kivlichan}, \citenamefont {McClean}, \citenamefont {Wiebe}, \citenamefont
  {Gidney}, \citenamefont {Aspuru-Guzik}, \citenamefont {Chan},\ and\
  \citenamefont {Babbush}}]{KivLinearDepth}%
  \BibitemOpen
  \bibfield  {author} {\bibinfo {author} {\bibfnamefont {I.~D.}\ \bibnamefont
  {Kivlichan}}, \bibinfo {author} {\bibfnamefont {J.}~\bibnamefont {McClean}},
  \bibinfo {author} {\bibfnamefont {N.}~\bibnamefont {Wiebe}}, \bibinfo
  {author} {\bibfnamefont {C.}~\bibnamefont {Gidney}}, \bibinfo {author}
  {\bibfnamefont {A.}~\bibnamefont {Aspuru-Guzik}}, \bibinfo {author}
  {\bibfnamefont {G.~K.-L.}\ \bibnamefont {Chan}}, \ and\ \bibinfo {author}
  {\bibfnamefont {R.}~\bibnamefont {Babbush}},\ }\href {\doibase
  10.1103/PhysRevLett.120.110501} {\bibfield  {journal} {\bibinfo  {journal}
  {Phys. Rev. Lett.}\ }\textbf {\bibinfo {volume} {120}},\ \bibinfo {pages}
  {110501} (\bibinfo {year} {2018})}\BibitemShut {NoStop}%
\bibitem [{\citenamefont {Jiang}\ \emph {et~al.}(2018)\citenamefont {Jiang},
  \citenamefont {Sung}, \citenamefont {Kechedzhi}, \citenamefont
  {Smelyanskiy},\ and\ \citenamefont {Boixo}}]{jiang2018hubbard}%
  \BibitemOpen
  \bibfield  {author} {\bibinfo {author} {\bibfnamefont {Z.}~\bibnamefont
  {Jiang}}, \bibinfo {author} {\bibfnamefont {K.~J.}\ \bibnamefont {Sung}},
  \bibinfo {author} {\bibfnamefont {K.}~\bibnamefont {Kechedzhi}}, \bibinfo
  {author} {\bibfnamefont {V.~N.}\ \bibnamefont {Smelyanskiy}}, \ and\ \bibinfo
  {author} {\bibfnamefont {S.}~\bibnamefont {Boixo}},\ }\href {\doibase
  10.1103/PhysRevApplied.9.044036} {\bibfield  {journal} {\bibinfo  {journal}
  {Phys. Rev. Applied}\ }\textbf {\bibinfo {volume} {9}},\ \bibinfo {pages}
  {044036} (\bibinfo {year} {2018})}\BibitemShut {NoStop}%
\bibitem [{\citenamefont {Nielsen}\ and\ \citenamefont
  {Chuang}(2002)}]{nielsen2002quantum}%
  \BibitemOpen
  \bibfield  {author} {\bibinfo {author} {\bibfnamefont {M.~A.}\ \bibnamefont
  {Nielsen}}\ and\ \bibinfo {author} {\bibfnamefont {I.}~\bibnamefont
  {Chuang}},\ }\href@noop {} {\enquote {\bibinfo {title} {Quantum computation
  and quantum information},}\ } (\bibinfo {year} {2002})\BibitemShut {NoStop}%
\bibitem [{\citenamefont {Higgott}\ \emph {et~al.}(2019)\citenamefont
  {Higgott}, \citenamefont {Wang},\ and\ \citenamefont
  {Brierley}}]{higgott2018variational}%
  \BibitemOpen
  \bibfield  {author} {\bibinfo {author} {\bibfnamefont {O.}~\bibnamefont
  {Higgott}}, \bibinfo {author} {\bibfnamefont {D.}~\bibnamefont {Wang}}, \
  and\ \bibinfo {author} {\bibfnamefont {S.}~\bibnamefont {Brierley}},\ }\href
  {\doibase 10.22331/q-2019-07-01-156} {\bibfield  {journal} {\bibinfo
  {journal} {{Quantum}}\ }\textbf {\bibinfo {volume} {3}},\ \bibinfo {pages}
  {156} (\bibinfo {year} {2019})}\BibitemShut {NoStop}%
\bibitem [{\citenamefont {McClean}\ \emph
  {et~al.}(2017{\natexlab{b}})\citenamefont {McClean}, \citenamefont
  {Kimchi-Schwartz}, \citenamefont {Carter},\ and\ \citenamefont
  {de~Jong}}]{PhysRevA.95.042308}%
  \BibitemOpen
  \bibfield  {author} {\bibinfo {author} {\bibfnamefont {J.~R.}\ \bibnamefont
  {McClean}}, \bibinfo {author} {\bibfnamefont {M.~E.}\ \bibnamefont
  {Kimchi-Schwartz}}, \bibinfo {author} {\bibfnamefont {J.}~\bibnamefont
  {Carter}}, \ and\ \bibinfo {author} {\bibfnamefont {W.~A.}\ \bibnamefont
  {de~Jong}},\ }\href {\doibase 10.1103/PhysRevA.95.042308} {\bibfield
  {journal} {\bibinfo  {journal} {Phys. Rev. A}\ }\textbf {\bibinfo {volume}
  {95}},\ \bibinfo {pages} {042308} (\bibinfo {year}
  {2017}{\natexlab{b}})}\BibitemShut {NoStop}%
\bibitem [{\citenamefont {Colless}\ \emph {et~al.}(2018)\citenamefont
  {Colless}, \citenamefont {Ramasesh}, \citenamefont {Dahlen}, \citenamefont
  {Blok}, \citenamefont {Kimchi-Schwartz}, \citenamefont {McClean},
  \citenamefont {Carter}, \citenamefont {de~Jong},\ and\ \citenamefont
  {Siddiqi}}]{PhysRevX.8.011021}%
  \BibitemOpen
  \bibfield  {author} {\bibinfo {author} {\bibfnamefont {J.~I.}\ \bibnamefont
  {Colless}}, \bibinfo {author} {\bibfnamefont {V.~V.}\ \bibnamefont
  {Ramasesh}}, \bibinfo {author} {\bibfnamefont {D.}~\bibnamefont {Dahlen}},
  \bibinfo {author} {\bibfnamefont {M.~S.}\ \bibnamefont {Blok}}, \bibinfo
  {author} {\bibfnamefont {M.~E.}\ \bibnamefont {Kimchi-Schwartz}}, \bibinfo
  {author} {\bibfnamefont {J.~R.}\ \bibnamefont {McClean}}, \bibinfo {author}
  {\bibfnamefont {J.}~\bibnamefont {Carter}}, \bibinfo {author} {\bibfnamefont
  {W.~A.}\ \bibnamefont {de~Jong}}, \ and\ \bibinfo {author} {\bibfnamefont
  {I.}~\bibnamefont {Siddiqi}},\ }\href {\doibase 10.1103/PhysRevX.8.011021}
  {\bibfield  {journal} {\bibinfo  {journal} {Phys. Rev. X}\ }\textbf {\bibinfo
  {volume} {8}},\ \bibinfo {pages} {011021} (\bibinfo {year}
  {2018})}\BibitemShut {NoStop}%
\bibitem [{\citenamefont {Neuscamman}\ \emph {et~al.}(2011)\citenamefont
  {Neuscamman}, \citenamefont {Changlani}, \citenamefont {Kinder},\ and\
  \citenamefont {Chan}}]{neuscamman2011TC_transform}%
  \BibitemOpen
  \bibfield  {author} {\bibinfo {author} {\bibfnamefont {E.}~\bibnamefont
  {Neuscamman}}, \bibinfo {author} {\bibfnamefont {H.}~\bibnamefont
  {Changlani}}, \bibinfo {author} {\bibfnamefont {J.}~\bibnamefont {Kinder}}, \
  and\ \bibinfo {author} {\bibfnamefont {G.~K.-L.}\ \bibnamefont {Chan}},\
  }\href {\doibase 10.1103/PhysRevB.84.205132} {\bibfield  {journal} {\bibinfo
  {journal} {Phys. Rev. B}\ }\textbf {\bibinfo {volume} {84}},\ \bibinfo
  {pages} {205132} (\bibinfo {year} {2011})}\BibitemShut {NoStop}%
\bibitem [{\citenamefont {Wahlen-Strothman}\ \emph {et~al.}(2015)\citenamefont
  {Wahlen-Strothman}, \citenamefont {Jim\'enez-Hoyos}, \citenamefont
  {Henderson},\ and\ \citenamefont {Scuseria}}]{Wahlen2015TC_transforms}%
  \BibitemOpen
  \bibfield  {author} {\bibinfo {author} {\bibfnamefont {J.~M.}\ \bibnamefont
  {Wahlen-Strothman}}, \bibinfo {author} {\bibfnamefont {C.~A.}\ \bibnamefont
  {Jim\'enez-Hoyos}}, \bibinfo {author} {\bibfnamefont {T.~M.}\ \bibnamefont
  {Henderson}}, \ and\ \bibinfo {author} {\bibfnamefont {G.~E.}\ \bibnamefont
  {Scuseria}},\ }\href {\doibase 10.1103/PhysRevB.91.041114} {\bibfield
  {journal} {\bibinfo  {journal} {Phys. Rev. B}\ }\textbf {\bibinfo {volume}
  {91}},\ \bibinfo {pages} {041114} (\bibinfo {year} {2015})}\BibitemShut
  {NoStop}%
\end{thebibliography}%

\onecolumngrid
\section{Appendix}\label{Sec:Appendix}
Here, we derive the evolution of parameters formula given in Eq.~(\ref{Eq:Imag_time_Matrix_vec}). Our derivation closely follows that given in Ref.~\cite{mcardle2018variational}. McLachlan's variational principle applied to the modified imaginary time Schr\"odinger equation, is given by
 \begin{equation}
 	\delta \|({\partial}/{\partial \tau} + H'-\Re(E_\tau))\ket{\phi(\tau)}\|=0
 \end{equation}
where
\begin{equation}
\|({\partial}/{\partial \tau} + H'-\Re(E_\tau))\ket{\phi(\tau)}\|=\bigg{(}({\partial}/{\partial \tau} + H'-E_R)\ket{\phi(\tau)}\bigg{)}^\dag \bigg{(} {\partial}/{\partial \tau} + H'-E_R)\ket{\phi(\tau)}\bigg{)},
\end{equation}
and $E_R = \Re(\braket{{\phi(\tau)}|H'|{\phi(\tau)}})$.
For a general quantum state, McLachlan's variational principle recovers the imaginary time evolution
\begin{equation}
    \frac{\partial \ket{\phi(\tau)}}{\partial \tau} = -[H' - \Re(E_\tau)]\ket{\phi(\tau)}.
\end{equation}

We restrict the algorithm to states that can be created by the ansatz $\ket{\Phi(\tau)}=\ket{\Phi(\theta_1,\theta_2,\dots,\theta_N)}$. We can project the imaginary time evolution onto this subspace using McLachlan's variational principle. Replacing $\ket{\phi(\tau)}$ with $\ket{\Phi(\tau)}$, yields
\begin{equation}
	\begin{aligned}
		\|({\partial}/{\partial \tau} + H'-E_R)\ket{\Phi(\tau)}\|=&\bigg{(}({\partial}/{\partial \tau} + H'-E_R)\ket{\Phi(\tau)}\bigg{)}^\dag\bigg{(}({\partial}/{\partial \tau} + H'-E_R)\ket{\Phi(\tau)}\bigg{)},\\
		=&\bigg{(}\frac{\partial \bra{\Phi}}{\partial \tau } + \bra{\Phi}H'^\dag - \bra{\Phi} E_R\bigg{)}\bigg{(}\frac{\partial \ket{\Phi}}{\partial \tau} + H'\ket{\Phi} - E_R \ket{\Phi}\bigg{)}  \\
		=&\sum_{i,j}\frac{\partial \bra{\Phi(\tau)}}{\partial \theta_i}\frac{\partial \ket{\Phi(\tau)}}{\partial \theta_j}\dot{\theta}_i \dot{\theta}_j+ \sum_{i}\frac{\partial \bra{\Phi(\tau)}}{\partial \theta_i}(H'-E_R)\ket{\Phi(\tau)}\dot{\theta}_i\\
		+& \sum_{i}\bra{\Phi(\tau)}(H'^\dag-E_R)\frac{\partial \ket{\Phi(\tau)}}{\partial \theta_i}\dot{\theta}_i +\bra{\Phi(\tau)}	(H'^\dag-E_R)(H'-E_R)\ket{\Phi(\tau)}. 
\end{aligned}
\end{equation}
Focusing on $\dot{\theta}_i$, we obtain 
\begin{equation}
	\begin{aligned}
		\frac{\partial \|({\partial}/{\partial \tau} + H'-E_R)\ket{\Phi(\tau)}\|}{\partial \dot{\theta}_i}&=\sum_{j}\left(\frac{\partial \bra{\Phi(\tau)}}{\partial \theta_i}\frac{\partial \ket{\Phi(\tau)}}{\partial \theta_j}+\frac{\partial \bra{\Phi(\tau)}}{\partial \theta_j}\frac{\partial \ket{\Phi(\tau)}}{\partial \theta_i}\right)\dot{\theta}_j \\
		&+\frac{\partial \bra{\Phi(\tau)}}{\partial \theta_i}(H'-E_R)\ket{\Phi(\tau)}+ \bra{\Phi(\tau)}(H'^\dag-E_R)\frac{\partial \ket{\Phi(\tau)}}{\partial \theta_i}.
\end{aligned}
\end{equation}

McLachlan's variational principle requires 
\begin{equation}
	\frac{\partial \|({\partial}/{\partial \tau} + H'-E_R)\ket{\Phi(\tau)}\|}{\partial \dot{\theta}_j} = 0,
\end{equation}

We then let 
\begin{equation}
\begin{aligned}
	A_{ij} &= \Re\left(\frac{\partial \bra{\Phi(\tau)}}{\partial \theta_i}\frac{\partial \ket{\Phi(\tau)}}{\partial \theta_j}\right),\\
	C_i &= \Re\left(\frac{\partial \bra{\Phi(\tau)}}{\partial \theta_i} H'\ket{\Phi(\tau)}\right),
\end{aligned}
\end{equation}
and use 
\begin{equation}
    -E_R\bigg{(}\frac{\partial \bra{\Phi(\tau)}}{\partial \theta_i} \ket{\Phi(\tau)} + \bra{\Phi(\tau)} \frac{\partial \ket{\Phi(\tau)}}{\partial \theta_i} \bigg{)} = -E_R\bigg{(}\frac{\partial}{\partial \theta_i} \braket{\Phi(\tau) | \Phi(\tau)}\bigg{)} = -E_R\bigg{(}\frac{\partial}{\partial \theta_i} 1\bigg{)} = 0
\end{equation}
to obtain an equation for the evolution of the parameters $\vec{\theta}$ under imaginary time evolution

\begin{equation}
	\begin{aligned}
		\sum_jA_{ij}\dot{\theta}_j = - C_i.
	\end{aligned}
\end{equation}
This is identical to the equation derived in Ref.~\cite{mcardle2018variational}.

\end{document}